\documentclass[sn-mathphys]{sn-jnl}
\usepackage{graphicx}
\usepackage{amsmath,amssymb}
\usepackage{url}

\newcommand{\be}{\begin{equation}}
\newcommand{\ee}{\end{equation}}
\newcommand{\ba}{\begin{array}}
\newcommand{\ea}{\end{array}}
\newcommand{\bea}{\begin{eqnarray}}
\newcommand{\eea}{\end{eqnarray}}

\def\beq{\begin{equation}}
\def\eeq#1{\label{#1}\end{equation}}
\def\eeqn{\end{equation}}

\newenvironment{Eqnarray}%
   {\arraycolsep 0.14em\begin{eqnarray}}{\end{eqnarray}}
\def\beqa{\begin{Eqnarray}}
\def\eeqa#1{\label{#1}\end{Eqnarray}}
\def\eeqan{\end{Eqnarray}}

\let\bar=\overbar

\def\lsim{\mathrel{\raise.3ex\hbox{$<$\kern-.75em\lower1ex\hbox{$\sim$}}}}

\def\del{\partial}
\def\Dslash{\not{\hbox{\kern-4pt $D$}}}
\def\dslash{\not{\hbox{\kern-2pt $\del$}}}
\def\pslash{\not{\hbox{\kern-2pt $p$}}}
\def\ETmiss{\not{\hbox{\kern-4pt $E$}}_T}

\def\Dlr{\mathrel{\raise1.5ex\hbox{$\leftrightarrow$\kern-1em\lower1.5ex\hbox{$D$}}}}

\def\ee{e^+e^-}

\def\MSB{{\bar{M \kern -2pt S}}}
\def\msb{{\bar{\scriptsize M \kern -1pt S}}}

\def\drb{{\bar{\scriptsize D \kern -1pt R}}}

\jyear{2021}%

\theoremstyle{thmstyleone}%
\theoremstyle{thmstyletwo}%

\theoremstyle{thmstylethree}%

\begin{document}

\title[Detection of Early-Universe GW Signatures and Fundamental Physics]{Detection of Early-Universe Gravitational-Wave Signatures and Fundamental Physics}


\author[1]{\fnm{Robert} \sur{Caldwell}}
\author[2]{\fnm{Yanou} \sur{Cui}}
\author[3]{\fnm{Huai-Ke} \sur{Guo}}
\author[4]{\fnm{Vuk} \sur{Mandic}}
\author[5]{\fnm{Alberto} \sur{Mariotti}} 
\author[6]{\fnm{Jose Miguel} \sur{No}}
\author[7]{\fnm{Michael J.} \sur{Ramsey-Musolf}}
\author[8]{\fnm{Mairi} \sur{Sakellariadou}}
\author[9]{\fnm{Kuver} \sur{Sinha}}
\author[10]{\fnm{Lian-Tao} \sur{Wang}}
\author[11]{\fnm{Graham} \sur{White${}^{\text{11}}$}}
\author[3]{\fnm{Yue} \sur{Zhao}}
\author[12,13,14]{\fnm{Haipeng} \sur{An}}
\author[15,14]{\fnm{Ligong} \sur{Bian}}
\author[16,17]{\fnm{Chiara} \sur{Caprini}}
\author[18]{\fnm{Sebastien} \sur{Clesse}}
\author[19]{\fnm{James M.} \sur{Cline}}
\author[16,20]{\fnm{Giulia} \sur{Cusin}}
\author[21]{\fnm{Bartosz} \sur{Fornal}}
\author[6]{\fnm{Ryusuke} \sur{Jinno}}
\author[19]{\fnm{Benoit} \sur{Laurent}}
\author[22]{\fnm{Noam} \sur{Levi${}^{\text{22}}$}}
\author[4]{\fnm{Kun-Feng} \sur{Lyu}}
\author[23]{\fnm{Mario} \sur{Martinez}}
\author[24]{\fnm{Andrew L.} \sur{Miller }}
\author[25]{\fnm{Diego} \sur{Redigolo}}
\author[4]{\fnm{Claudia} \sur{Scarlata}}
\author[5]{\fnm{Alexander} \sur{Sevrin}}
\author[3]{\fnm{Barmak Shams Es } \sur{Haghi}}
\author[26,27,28,29]{\fnm{Jing} \sur{Shu}}
\author[30]{\fnm{Xavier} \sur{Siemens}}
\author[31]{\fnm{Dani\`ele A.} \sur{Steer}}
\author[32]{\fnm{Raman} \sur{Sundrum}}
\author[33]{\fnm{Carlos} \sur{Tamarit${}^{\text{33}}$}}
\author[34]{\fnm{David J.} \sur{Weir}}
\author[35]{\fnm{Ke-Pan} \sur{Xie}}
\author[3]{\fnm{Feng-Wei} \sur{Yang}}
\author[36]{\fnm{Siyi} \sur{Zhou}}

\equalcont{These authors contributed equally to this work.}

\affil[1]{Department of Physics and Astronomy, Dartmouth College, Hanover, NH 03755, USA}
\affil[2]{Department of Physics and Astronomy, University of California, Riverside, CA 92521, USA}
\affil[3]{Department of Physics and Astronomy, University of Utah, Salt Lake City, UT, 84112, USA}
\affil[4]{School of Physics and Astronomy, University of Minnesota, Minneapolis, MN 55455, USA}
\affil[5]{Theoretische Natuurkunde and IIHE/ELEM, Vrije Universiteit Brussel, and International Solvay Institutes, Pleinlaan 2, 1050 Brussels, Belgium}
\affil[6]{Instituto de F{\'{\i}}sica Te{\'o}rica UAM/CSIC, C/ Nicol{\'a}s Cabrera 13-15, Campus de Cantoblanco, 28049, Madrid, Spain}
\affil[7]{Tsung Dao Lee Institute/Shanghai Jiao Tong University, Shanghai 200120 China, and University of Massachusetts, Amherst, MA 01003 USA}
\affil[8]{Physics Department, King's College London, Strand, London WC2R 2LS, UK}
\affil[9]{Department of Physics and Astronomy, University of Oklahoma, Norman, OK 73019, USA}
\affil[10]{Department of Physics, University of Chicago, Chicago, IL 60637, USA}
\affil[11]{Kavli IPMU (WPI), UTIAS, The University of Tokyo, Kashiwa, Chiba 277-8583, Japan}
\affil[12]{Department of Physics, Tsinghua University, Beijing 100084, China}
\affil[13]{Center for High Energy Physics, Tsinghua University, Beijing 100084, China}
\affil[14]{Center for High Energy Physics, Peking University, Beijing, 100871, China}
\affil[15]{Department of Physics and Chongqing Key Laboratory for Strongly Coupled Physics, Chongqing University, Chongqing 401331, P. R. China}

\affil[16]{Theoretical Physics Department, University of Geneva, 1211 Geneva, Switzerland}
\affil[17]{CERN, Theoretical Physics Department, 1 Esplanade des Particules, CH-1211 Gen\`eve 23, Switzerland}
\affil[18]{Service de Physique Th{\'e}orique (CP225), University of Brussels (ULB), Boulevard du Triomphe, 1050, Brussels, Belgium}
\affil[19]{Department of Physics, McGill University, Montr{\'e}al, Qu{\'e}bec, H3A2T8, Canada}
\affil[20]{Sorbonne Universit{\'e}, CNRS, UMR 7095, Institut d'Astrophysique de Paris, 75014 Paris, France}
\affil[21]{Department of Chemistry and Physics, Barry University, Miami Shores, FL 33161, USA}
\affil[22]{Raymond and Beverly Sackler School of Physics and Astronomy, Tel-Aviv University, Tel-Aviv 69978, Israel}
\affil[23]{Institut de F{\'{\i}}sica d'Altes Energies, Barcelona Institute of Science and Technology and ICREA, E-08193 Barcelona, Spain}
\affil[24]{Universit{\'e} catholique de Louvain, 1348 Louvain-la-Neuve, Belgium}
\affil[25]{INFN, Sezione di Firenze Via G. Sansone 1, 50019 Sesto Fiorentino, Italy}
\affil[26]{CAS Key Laboratory of Theoretical Physics, Insitute of Theoretical
Physics, Chinese Academy of Sciences, Beijing 100190, P.R.China}
\affil[27]{School of Physical Sciences, University of Chinese Academy of Sciences, Beijing 100049, China} 
\affil[28]{School of Fundamental Physics and Mathematical Sciences, Hangzhou Institute for Advanced Study, University of Chinese Academy of Sciences, Hangzhou 310024, China}
\affil[29]{International Center for Theoretical Physics Asia-Pacific, Beijing/Hanzhou, China}
\affil[30]{Department of Physics, Oregon State University, Corvallis, OR 97331, USA}
\affil[31]{Laboratoire Astroparticule et Cosmologie, CNRS, Universit\'e Paris Cit\'e, 75013 Paris France}
\affil[32]{University of Maryland, College Park, MD 20742, USA}
\affil[33]{Physik-Department T70, Technische Universit\"at M\"unchen, James-Franck-Stra{\ss}e, 85748 Garching, Germany}
\affil[34]{Department of Physics and Helsinki Institute of Physics, P.O. Box 64, 00014 University of Helsinki, Finland}
\affil[35]{Department of Physics and Astronomy, University of Nebraska, Lincoln, NE 68588, USA}
\affil[36]{Department of Physics, Kobe University, Kobe 657-8501, Japan}


\abstract{
\noindent Detection of a gravitational-wave signal of non-astrophysical origin would be a landmark discovery, potentially providing a significant clue to some of our most basic, big-picture scientific questions about the Universe. In this white paper, we survey the leading early-Universe mechanisms that may produce a detectable signal -- including inflation, phase transitions, topological defects, as well as primordial black holes -- and highlight the connections to fundamental physics. We review the complementarity with collider searches for new physics, and multimessenger probes of the large-scale structure of the Universe.
}




\maketitle

\newpage

\tableofcontents

\section{Introduction}

Despite their remarkable successes, both the Standard Model of Particle Physics and the Standard Model of Cosmology face multiple open questions. Examples include the origin and composition of dark matter, the origin of dark energy, the evolution of the Universe during the first minute after the Big Bang (including the inflationary phase as well as possible other phase transitions), the particle physics at the TeV and higher energy scales, the mechanism of electroweak symmetry breaking, and others. Many of these phenomena could have left measurable imprints in the form of gravitational waves (GWs) coming from the early Universe, providing a unique connection between GW physics and fundamental physics. Given the very high energies associated with many of these phenomena, often unachievable in laboratories and accelerators, GWs may provide the only experimental handle to probe these domains for new physics. (Snowmass white paper ``Future Gravitational-Wave Detector Facilities'' \cite{Ballmer:2022uxx} gives an extensive survey of planned and proposed GW observatories. Snowmass white paper ``Spacetime Symmetries and Gravitational Physics''~\cite{Adelberger:2022sve} provides
an overview of high-sensitivity small experiments that can be used for GW detections.) 
 
The connection between high energy physics, cosmology, and GW physics has been investigated through many facets, and can be illustrated using different perspectives. On one hand, there are open questions that are inevitably linked with the cosmology of our Universe and that may be partially decoupled from the Standard Model of Particle Physics. We dedicate two individual Sections to such questions. Section \ref{sec:earlyphases} is dedicated to the questions about the origin of our Universe. This includes the inflationary and other early phases in the evolution of the Universe that could have generated a stochastic GW background (SGWB). (For synergy, see Snowmass white papers on inflation \cite{Achucarro:2022qrl} and high energy physics model-building of the early Universe \cite{Flauger:2022hie,Asadi:2022njl}.) Later, Section \ref{sec:dm} is dedicated to the open problem of dark matter. (Many of the open problems in cosmology are presented in the Snowmass white paper ``Cosmology Intertwined: A Review of the Particle Physics, Astrophysics, and Cosmology Associated with the Cosmological Tensions and Anomalies'' \cite{Abdalla:2022yfr}.) Since there are many possible strategies to address the dark matter problem, the dark matter mechanisms that generate GW signals are necessarily diverse. For instance, a dark matter particle could leave an imprint in GW signals by distorting the binary merger dynamics, while dark matter in the form of primordial black holes would provide a new source of binary mergers GW signals. (For synergy, see the Snowmass white paper ``Observational Facilities to Study Dark Matter''~\cite{Chakrabarti:2022cbu}.
See also the Snowmass white paper ``Primordial Black Hole Dark Matter'' \cite{Bird:2022wvk} for a detailed discussion of the implications of primordial black holes. There are also significant overlaps with the Snowmass white papers ``Astrophysical and Cosmological Probes of Dark Matter" \cite{Boddy:2022knd} and Snowmass white paper: ``Cosmic Probes of Fundamental Physics Probing dark matter with small-scale astrophysical observations'' \cite{Snowmass2021:CosmoDM_SmallScale}.)

Many fundamental questions in particle physics  -- the electroweak hierarchy problem and the mechanism behind electroweak symmetry breaking, the strong CP problem, the matter-antimatter asymmetry, neutrino masses, to some extent also dark matter -- are typically addressed by introducing new particles and symmetries to the Standard Model (SM), with new, and possibly dark, sectors. These new symmetries may break through phase transitions during the evolution of the Universe, possibly constituting important new sources of a SGWB. GW production mechanisms include both the dynamics of the phase transition if it is first order, as discussed in Section \ref{section_PT}, or  topological defects if they are created during the phase transition, as covered in Section \ref{sec:defects}. 

Finally, there is a great potential for multimessenger complementarity between GW observations and the standard techniques for probing cosmology and particle physics. In Section \ref{sec:gwcollider}, we explore such complementarity between GW observations and the future collider experiments in the context of probing TeV-scale physics. This includes studies of the electroweak symmetry breaking in colliders and associated GWs from phase transitions in the early Universe. It also includes other possible symmetries and associated phase transitions, such as those related to supersymmetry. (For synergy with collider-based probes, see the Snowmass white paper ``Probing the Electroweak Phase Transition with Exotic Higgs Decays'' \cite{Carena:2022yvx}. Further GW probes of fundamental physics are described in the Snowmass white paper ``Fundamental Physics and Beyond the Standard Model''  \cite{Berti:2022wzk}.) In Section \ref{sec:gwem} we explore complementarity between GW observations and traditional electromagnetic observations of the large-scale structure, such as the cosmic microwave background (CMB) or weak gravitational lensing. Directional correlations between these observations, codified in angular cross-correlation spectra, may offer unique information about the early phases in the evolution of the Universe and perhaps shed additional information on the dark matter problem. (Snowmass white papers ``Cosmology and Fundamental Physics from the Three-Dimensional Large Scale Structure'' \cite{Ferraro:2022cmj} and ``Cosmic Microwave Background Measurements'' \cite{Chang:2022tzj} provide a deeper study of the connections of large scale structure and CMB measurements to fundamental physics.) We offer concluding remarks in Section \ref{sec:conclusion}.

We note that SGWB could also be generated by astrophysical sources such as mergers of binary black hole and binary neutron star systems. This astrophysical SGWB can act as a foreground to the cosmological (early universe) SGWB. Suppression or removal of the astrophysical stochastic GW foreground is an active area of research, with multiple techniques being explored \cite{PhysRevLett.118.151105,PhysRevD.102.024051,PhysRevD.102.063009,PhysRevLett.125.241101}. These efforts have only had partial success to date and further studies in this direction are needed. We will not discuss this problem any further in this paper.

\begin{figure}[!t]
\begin{center}
\includegraphics[width=1\hsize]{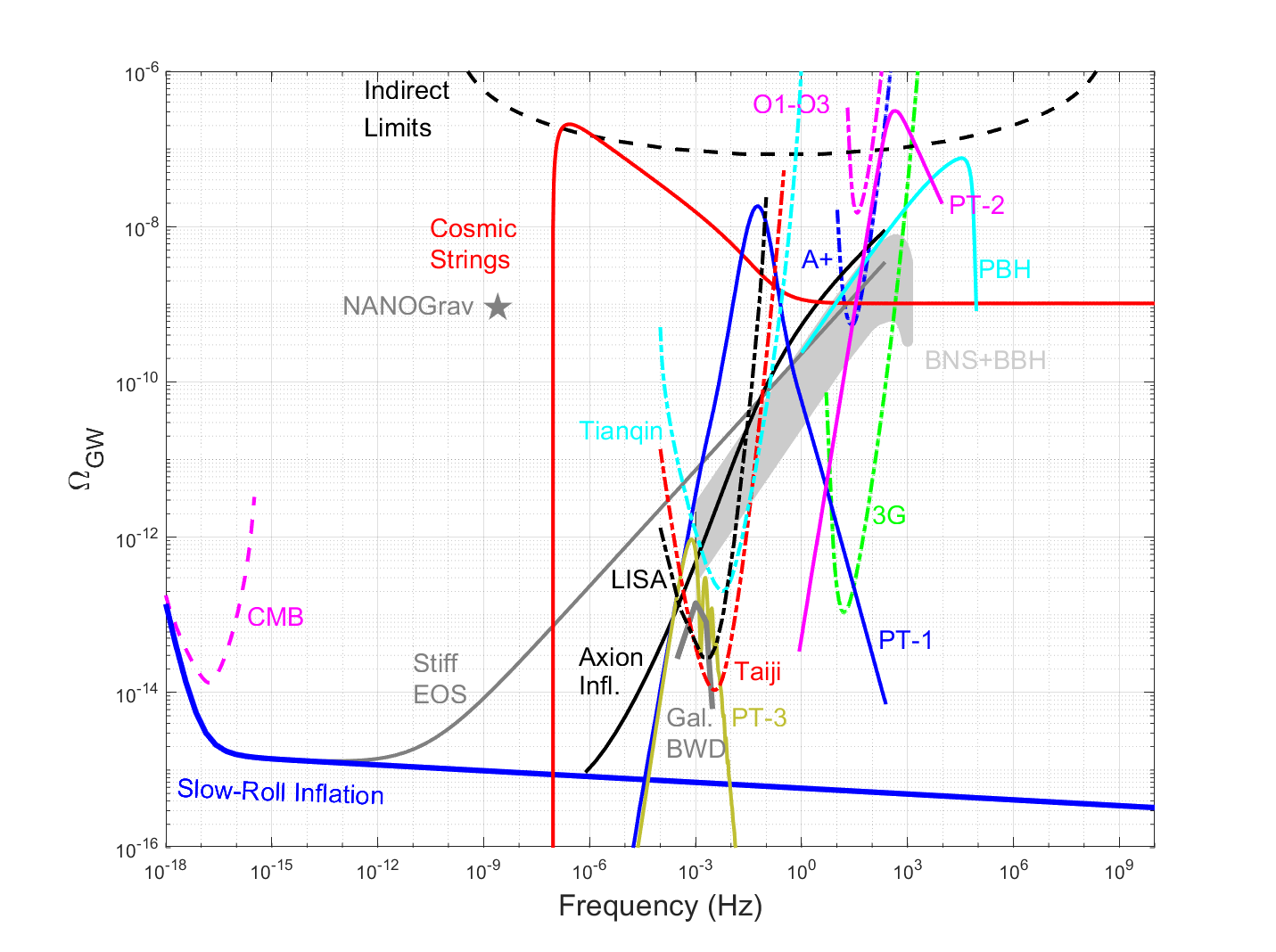}
\end{center}
\caption{Landscape of gravitational wave cosmology. Experimental results include: O1-O3 LIGO-Virgo upper limits~\cite{PhysRevD.104.022004}, indirect limits from big bang nucleosynthesis~\cite{PhysRevX.6.011035}, CMB limits~\cite{PhysRevX.6.011035}, and NANOGrav pulsar timing measurement~\cite{Arzoumanian_2020}, as well as projected sensitivities of the third generation (3G) terrestrial GW detectors~\cite{EinsteinTelescope,CosmicExplorer} and space-borne LISA~\cite{LISA}, Taiji~\cite{Hu:2017mde}, and Tianqin~\cite{TianQin:2015yph}. Theoretical models include examples of slow-roll inflation~\cite{turner}, first-order phase transitions (PT-1~\cite{Ellis:2019oqb}, PT-2~\cite{DelleRose:2019pgi}, and PT-3~\cite{An:2022}), Axion Inflation~\cite{peloso_parviol}, Primordial Black Hole model \cite{Wang:2016ana}, hypothetical stiff equation of state in the early universe~\cite{boylebuonanno}, and foregrounds due to binary black hole/neutron stars~\cite{PhysRevD.104.022004} and galactic binary white dwarfs~\cite{LISA}.
}
\label{fig:sample}
\end{figure}

\section{Early Phases in the Evolution of the Universe\label{sec:earlyphases}}

    
    
    
    


A variety of physical processes in the early Universe may generate GWs.
As the universe expands and the temperature drops, phase transitions may take place followed by spontaneously broken symmetries.

First order phase transitions may generate GWs \cite{PhysRevLett.69.2026} within the frequency range of present \cite{Romero:2021kby} or future \cite{Caprini:2019egz} interferometers, providing a way to test particle physics models beyond the SM. 
Phase transitions followed by spontaneous symmetry breaking may lead to the production of topological defects as relics of the previous more symmetric phase of the Universe; they are  characterised by the homotopy group of the false vacuum. One-dimensional topological defects, called cosmic strings \cite{Vilenkin:2000jqa,Kibble:1976sj,Sakellariadou:2006qs}, were shown \cite{Jeannerot:2003qv} to be generically formed at the end of hybrid inflation within the context of grand unified theories. 
Cosmic strings, analogues of vortices in condensed matter systems, 
 leave several observational signatures, opening up a new window to fundamental physics at energy scales much above the ones reached by accelerators. 
The production of GWs by cosmic strings \cite{PhysRevD.31.3052,Sakellariadou:1990ne} is one of the most promising observational signatures; they can be accessed by interferometers \cite{LIGOScientific:2021nrg,Auclair:2019wcv,Boileau:2021gbr}, as illustrated in Fig. \ref{fig:sample}. Domain walls, \cite{Vilenkin:1981zs,Sakellariadou:1990ne,Sakellariadou:1991sd,Gleiser:1998na,Hiramatsu:2010yz,Dunsky:2021tih}, textures \cite{Fenu:2009qf}, and indirectly monopoles \cite{Dunsky:2021tih} can also source GWs; they are hitherto less studied than cosmic strings. 
A cosmic string network is mainly characterised by the string tension $G\mu (c = 1)$, where $G$ is Newton's constant and $\mu$ the mass per unit length. Observational data constrain the $G\mu$ parameter, which is related to the energy scale of the phase transition leading to cosmic string formation, and therefore it may also be related to the energy scale of inflation. Cosmic superstrings, predicted \cite{Sarangi:2002yt,Jones:2002cv} in superstring inspired inflationary models with spacetime-wrapping D-branes, are coherent macroscopic states of fundamental superstrings and D-branes extended in one macroscopic direction.  For cosmic superstrings, there is an additional parameter, namely the intercommutation probability. 
The dynamics of either a cosmic string or a cosmic superstring network is driven by the formation of loops and the emission of GWs.

Following the inflationary paradigm, vacuum fluctuations of the inflaton field may generate a scale-invariant spectrum of GWs  imprinted on the CMB B-mode polarization. Inflationary models generally predict very small values of the ratio of the power spectra of the tensor and scalar modes $(r\ll 1)$.  Combining Planck with BICEP2/Keck 2015 data yields an upper limit of $r<0.044$ \cite{Tristram:2020wbi}.
 Inflationary scenarios in which the inflaton couples to a gauge field may predict strongly blue-tilted GW spectra that may be consistent with CMB bounds at long wavelengths but within reach of direct detection experiments at short wavelengths. (See Ref.~\cite{Thorne:2017jft} for examples, or the forthcoming ``Cosmology with the Laser Interferometer Space Antenna,'' by the LISA Cosmology Working Group \cite{Auclair:2022lcg}, for further discussion.) Effective Field Theory inflationary models that do not respect the null-energy condition are also characterized by a blue tensor tilt \cite{Capurri:2020qgz}, as well as inflationary models with violation of slow roll \cite{Wang:2014kqa}, and potentially inflationary models within modified gravity. Finally, there are inflationary models that predict features in the stochastic GW background, for instance as the result of particle production during inflation \cite{Fumagalli:2020nvq}.

Despite its success, inflation does face drawbacks \cite{Goldwirth:1989pr,Calzetta:1992gv,Calzetta:1992bp,Borde:2001nh,Ijjas:2014nta,Brandenberger:2000as}, hence various alternatives have been proposed and worked out. Within bouncing cosmologies \cite{Brandenberger:2016vhg},  the pre-Big Bang \cite{Gasperini:1992em} or ekpyrotic model \cite{Khoury:2001wf}, the string gas cosmology \cite{Brandenberger:1988aj,Battefeld:2014uga}
 and the matter bounce scenario \cite{Finelli:2001sr,Brandenberger:2012zb}, 
have been long discussed in the literature. 

 The pre-Big Bang scenario \cite{Gasperini:1992em,Gasperini:2002bn} assumes that
gravity and cosmology are based on some particular version of superstring theory. According to this model, the Universe  emerges from a highly perturbative initial state preceding the Big Bang. 
In string theory, the dilaton is an additional massless mode, which while at late times may be assumed fixed, is dynamical in the very early Universe. Hence, the massless sector of string theory to which the graviton belongs is given by dilaton gravity. 
This model leads to the production of an amplified quasi-thermal spectrum of gravitons during the dilaton-driven phase. While the slope of the GW spectrum may change for modes crossing the horizon during the subsequent string phase, it remains characterized by an enhanced production of high frequency gravitons, irrespective of the particular value of the spectral index \cite{Brustein:1995ah}. For a wide region of the parameter space of the pre-Big Bang model, one can simultaneously generate a spectrum of scalar metric perturbations in agreement with Planck data and a stochastic background of primordial GWs within reach of the design sensitivity of aLIGO/Virgo and/or LISA \cite{Gasperini:2016gre}.

The ekpyrotic model \cite{Khoury:2001wf} is motivated by a specific realization of M-theory in which spacetime is 11-dimensional \cite{Horava:1995qa}. In this model, whereas the equation of motion of the cosmological perturbations depends on the potential of the scalar field, that of the GWs does not. The spectrum of GWs turns out to be very blue, implying that primordial GWs are negligible on scales of cosmological interest today. 

String gas cosmology is based on degrees of freedom and symmetries of superstring theory which are absent in an effective
field theory approach to early Universe cosmology.
This model has two fundamental elements. The first is the Hagedorn temperature defined as  the maximal temperature which a gas
of strings in thermal equilibrium can achieve. The second is the T-duality symmetry of the spectrum of string states.
This string gas cosmology model (tested also via numerical experiments \cite{Sakellariadou:1995vk})  leads to an almost scale-invariant spectrum of primordial GWs \cite{Brandenberger:2006vv}. However, whereas inflation typically leads to a red spectrum, string gas cosmology generically yields a slightly blue spectrum.

The matter bounce model \cite{Finelli:2001sr,Brandenberger:2012zb}  is based on the duality between the evolution of the canonical fluctuation variables in an exponentially expanding period and in a contracting phase with vanishing pressure. In this model, the amplitude of the GW spectrum is generically larger than that in  inflationary models ($r$ can be close to 1).


Having indicated the many possibilities during the expansion phase in the very early Universe, we now turn to the role that GWs play in the subsequent eras. GWs, because they free-stream through the entirety of cosmological history, provide us a way of probing cosmic history before Big Bang Nucleosynthesis (BBN).

$(i)$ Long-lasting GW sources as cosmic witnesses: The vast stretch of energies between the initial expansion phase of the Universe and BBN can accommodate a range of non-standard cosmological histories with equations of state parameter $w$ different from $w=1/3$. The GW spectrum of long-lasting sources such as inflation and  networks of cosmic strings spans a wide range of frequencies and has been studied  as cosmic witnesses \cite{DEramo:2019tit, Figueroa:2019paj}, especially to extract information about $w$.  Scenarios with $1/3 \leq w \leq 1$ 
develop a blue tilt for the tensor perturbations at large frequencies and certain portions of the parameter space may be detectable by LISA and/or terrestrial GW detectors \cite{boylebuonanno,Figueroa:2019paj,Giovannini_1,Giovannini_2}, c.f. Fig. \ref{fig:sample}.  Matter-dominated era with $w=0$ may cause kinks in the spectrum at frequencies corresponding to the onset of matter and radiation domination, which may be observable in the future at BBO and other experiments. 
If the  change in the equation of state is sudden enough, rapidly oscillating scalar perturbations may enhance the primordial GW spectrum from inflation; this effect has been pursued in the context of primordial black holes \cite{Inomata:2020lmk} and Q-balls \cite{White:2021hwi}.

$(ii)$ Phase transitions as cosmic witnesses: It is challenging to use shorter-lasting sources like phase transitions as a witness to  $w$ in the early universe, although some effort has been made in this direction \cite{Guo:2020grp, Hook:2020phx, Barenboim:2016mjm, Cai:2019cdl}. 
%
%
If the phase transition occurs during a phase dominated by an entity with a general equation of state $w$ that is not radiation, expressions for the parameters defining the phase transition should take into account the new dominant contribution to the energy density; if the phase transition is followed by an era dominated by an entity with state $w$, the modified redshifting changes the spectrum. In the deep infrared, super-horizon modes scale as $\propto k^{3+2\frac{3w-1}{3w+1}}$. 
A special case of phase transitions serving as a probe of inflationary physics is when the transition occurs \textit{during} inflation itself, possibly triggered by inflaton couplings to spectator fields. For such a source, the GW spectrum always contains a unique oscillatory feature, which can be used to identify the GW source~\cite{An:2020fff}.


A different avenue is to use GWs from phase transitions as a record of temperature anisotropies coming from a previous era, for example from inflation. Since the phase transition inherits primordial temperature anisotropies, it effectively serves as a copy of the CMB, but in an earlier, more pristine form \cite{Geller:2018mwu}. While this matching  is true for  adiabatic fluctuations, if the primordial fluctuations  carry an  isocurvature component then a richer scenario can emerge \cite{Kumar:2021ffi}. 

$(iii)$ GWs from inflationary reheating/preheating: The large  time-dependent field inhomogeneities that are characteristic of rapid particle production through parametric resonances during a preheating phase \cite{Kofman:1997yn, Kofman:1994rk, Amin:2014eta} can be a well-motivated source of GW production. Topics that have been explored in this context include  gauge preheating after axion inflation  \cite{Adshead:2019lbr, Adshead:2018doq}, self-resonance after single-field inflation and oscillon formation \cite{Zhou:2013tsa,Lozanov:2019ylm, Antusch:2016con, Amin:2018xfe, Hiramatsu:2020obh,Kou:2021bij}, as well as tachyonic preheating from a waterfall transition \cite{Garcia-Bellido:2007nns, Garcia-Bellido:2007fiu, Dufaux:2010cf}. The frequency of the resultant GW signal is typically too high, or the amplitude too small, to be detectable with near future GW detectors, although several ideas have been proposed recently \cite{Berlin:2021txa, Domcke:2022rgu}. (See also the Snowmass White Paper~\cite{Berlin:2022hfx}.)  The recent work of  \cite{Cui:2021are} demonstrates a promising GW detection prospect based on a preheating scenario in the framework of hybrid inflation, where a prolonged waterfall phase allows for an efficient transfer of energy from the scalar sector to an Abelian gauge field.  
For particular reheating mechanisms, the study of \cite{Haque:2021dha} has investigated the phases of early and late time reheating through imprints on primordial GWs.

\section{Phase Transitions}
\label{section_PT}

GWs from first order phase transitions (FOPTs) in the early Universe offer a unique way of probing particle physics models at energy scales otherwise inaccessible. The GW spectrum, with examples shown in Fig.~\ref{fig:sample}, is sensitive to the shape of the effective potential, which depends on the symmetry breaking pattern and the particle content of the theory. This provides access to regions of parameter
space unexplored so far in various extensions of the SM. GWs from a strong FOPT have a plethora of motivations in the early universe. For instance, new physics at the electroweak scale can lead to a strongly first order electroweak phase transition \cite{Ramsey-Musolf:2019lsf,Profumo:2007wc,Delaunay:2007wb,Huang:2016cjm,Chala:2018ari,Croon:2020cgk,Grojean:2006bp,Alves:2018jsw,Alves:2020bpi,Vaskonen:2016yiu,Dorsch:2016nrg,Chao:2017vrq,Wang:2019pet,Demidov:2017lzf,Ahriche:2018rao,Huang:2017rzf,Mohamadnejad:2019vzg,Baldes:2018nel,Huang:2018aja,Ellis:2019flb, Alves:2018oct, Alves:2019igs,Cline:2021iff,Chao:2021xqv,Liu:2021mhn,Zhang:2021alu,Cai:2022bcf,Liu:2022vkm} and large lepton asymmetries or different  quark masses can make the QCD transition strong \cite{Schwarz:2009ii,Middeldorf-Wygas:2020glx,Caprini:2010xv,vonHarling:2017yew}. Beyond this, a strong transition can occur in multistep phase transitions\footnote{See Refs. \cite{Weinberg:1974hy,Land:1992sm,Patel:2012pi,Patel:2013zla,Blinov:2015sna} for the viability of a multistep phase transition} \cite{Niemi:2018asa,Croon:2018new,Morais:2018uou,Morais:2019fnm,Angelescu:2018dkk,TripletGW2022}, B-L breaking \cite{Jinno:2016knw,Chao:2017ilw,Brdar:2018num,Okada:2018xdh,Marzo:2018nov,Bian:2019szo,Hasegawa:2019amx,Ellis:2019oqb,Okada:2020vvb} (or B/L breaking \cite{Fornal:2020esl}), flavour physics \cite{Greljo:2019xan,Fornal:2020ngq}, axions \cite{Dev:2019njv,VonHarling:2019rgb,DelleRose:2019pgi}, GUT symmetry breaking chains \cite{Hashino:2018zsi,Huang:2017laj,Croon:2018kqn,Brdar:2019fur,Huang:2020bbe}, supersymmetry breaking \cite{Fornal:2021ovz,Craig:2020jfv,Apreda:2001us,Bian:2017wfv}, hidden sector involving scalars \cite{Schwaller:2015tja,Baldes:2018emh,Breitbach:2018ddu,Croon:2018erz,Hall:2019ank,Baldes:2017rcu,Croon:2019rqu,Hall:2019rld,Hall:2019ank,Chao:2020adk,Dent:2022bcd}, neutrino mass models \cite{Li:2020eun,DiBari:2021dri,Zhou:2022mlz} and confinement \cite{Helmboldt:2019pan,Aoki:2019mlt,Helmboldt:2019pan,Croon:2019ugf,Croon:2019iuh,Garcia-Bellido:2021zgu,Huang:2020crf,Halverson:2020xpg,Kang:2021epo}.

Such phase transitions could occur at nearly any time during or after inflation. For example, the environment in which an electroweak-scale phase transition takes place, where bubbles expand in a plasma of relativistic SM particles, is very different to that prior to reheating. Much of what is discussed in this section relates in particular to what one might call \textsl{thermal phase transitions}, in which the bubbles typically nucleate thermally through the three-dimensional bounce action. Despite the name, not all thermal transitions efficiently transfer kinetic energy to the plasma, as will be discussed below. Nevertheless, as a class of scenarios they represent the most likely source of an observable SGWB due to a FOPT, and have the richest complementarity with other particle physics and cosmological observables.

A thermal phase transition begins with the nucleation of bubbles where the walls expand in a plasma of ultrarelativistic particles, and the interactions of the particles with the walls in large part determine the terminal wall velocity of the bubbles. 
GWs are first sourced by the colliding bubble walls \cite{Kosowsky:1991ua,Kosowsky:1992vn,Huber:2008hg,Jinno:2016vai} and the fluid shock configurations, detonation or deflagrations, accompanying them \cite{Kamionkowski:1993fg}, 
although for a thermal transition the shear stress in the walls and shocks is unlikely to be a substantial source of GWs \cite{Hindmarsh:2015qta}. 
For deflagrations, pressure may build up in front of the walls, deforming them and delaying completion of the transition through the formation of hot droplets \cite{Cutting:2019zws}. 
After the bubbles have merged, a bulk fluid velocity will remain in the plasma. 
At first the velocity perturbations will typically be longitudinal, unless the bubbles have been deformed during the transition due to hydrodynamic instabilities or deformations of the shape. 
In weak transitions, the fluid perturbation will take the form of longitudinal acoustic waves -- sound waves 
\cite{Hindmarsh:2013xza}. 
If the shock formation timescale $\tau_\text{sh} \sim L_*/v_\text{rms}$, where $L_*$ is a typical length scale in the fluid
linked to the mean bubble separation, 
and $v_\text{rms}$ is the root mean square fluid 3-velocity, is smaller than the Hubble time, shocks will form \cite{Caprini:2015zlo}. 
This is expected to occur for strong transitions, and to lead to turbulence \cite{Pen:2015qta}. 
%
Sound waves, acoustic, and vortical turbulence are all sources of GWs, lasting until the kinetic energy is dissipated by the plasma viscosity \cite{Caprini:2009yp}.
These different processes source GWs with different spectral shapes (see e.g.~\cite{Hindmarsh:2017gnf,Hindmarsh:2019phv,Niksa:2018ofa,Caprini:2009yp,RoperPol:2019wvy,Jinno:2020eqg}), which would allow us in principle to reconstruct the conditions in the Universe during and after a sufficiently strong FOPT.

The starting point in the calculation is the effective potential $V_{\rm eff}$ of a given model, consisting of three contributions: tree-level, one-loop Coleman-Weinberg, and finite temperature part.
The $V_{\rm eff}$ initially admits a vacuum at high temperature, typically at the origin of the field space, and starts to develop another one which becomes more and more energetically preferable as temperature drops.
Provided that the two vacua are separated by a potential barrier, the Universe then undergoes a FOPT to the lower-energy state. 
This is realized through nucleating bubbles of true vacuum, which then expand and collide with each other, eventually leaving the Universe in the
new vacuum state. 
Four parameters characterizing this picture dictate the resulting GWs: the nucleation temperature $T_*$, the bubble wall velocity $v_w$, the FOPT's strength $\alpha$ and its inverse duration $\beta$. We note the following issues and recent developments
regarding their calculations.

{\bf 1. Issues in perturbation theory.}
A perturbative treatement of the finite temperature potential is known to breakdown. The central problem is that the expansion parameter at finite temperature involves a mode occupation 
which diverges when the mass vanishes \cite{Linde:1980ts}. This can be partially addressed by resumming the most dangerous ``daisy'' diagrams. 
However, the usual resummation prescriptions have the issues that ($T$ dependent) UV divergences do not cancel at the same order \cite{Laine:2017hdk} and anyway does not address the issue of the slow convergence of perturbation theory. Including next to leading order corrections can change the predictions of the GW amplitude by many orders of magnitude \cite{Croon:2020cgk,Gould:2021oba}. 
At present, only the technique of dimensional reduction \cite{Kajantie:1995dw,Farakos:1994xh} performed at NLO using an $\hbar$ expansion provides a prescription to calculate thermodynamic parameters at $O(g^4)$ in a gauge independent way \cite{Croon:2020cgk,Gould:2021oba}. This method is challenging to use and has been applied to benchmarks in very few models. There are proposed alternatives to dimensional reduction \cite{Curtin:2016urg,Croon:2021vtc}. However, these are in need of development and testing and it is not yet obvious how to apply such techniques in a gauge invariant way.  Finally, for very weak transitions, the tachyonic mass of the physical Higgs is cancelled by thermal mass near the origin, leading to an unresolved infrared divergence which is probably captured by large differences in predictions of perturbation theory and Monte-Carlo simulations \cite{Gould:2019qek,Niemi:2020hto}. It is generally assumed that perturbation theory in its most sophisticated form should give accurate results provided a transition is strong enough, however this needs to be proven by careful comparison with Monte-Carlo simulations.

{\bf 2. Calculation of the bubble nucleation rate.}
An accurate evaluation of the nucleation rate $\Gamma_{\mathrm{nuc}}$ and its evolution with temperature is of paramount importance to defining the characteristic time scales of the transition. 
 For sufficiently fast transitions, $T_{\ast}$ and $\beta$ can be obtained by linearizing the rate  near $T_{\ast}$.
 This breaks down for slow transitions, which can be of great phenomenological interest, where the next order corrections must be accounted for~\cite{Ellis:2018mja}.
Various components contribute to $\Gamma_{\rm{nuc}}$, each meriting a separate discussion. Firstly, analytical solutions of the bounce EOM exist only for specific single field potentials~\cite{Fubini:1976jm, Coleman:1980aw,Duncan:1992ai, Adams:1993zs}, with progress made in the study of approximate single field potentials for light scalars~\cite{Dutta:2011rc, Aravind:2014aza,Espinosa:2018hue, Guada:2020ihz, Amariti:2020ntv}. Often, however, the underlying theory implies highly nonlinear equations of motion, or the existence of multiple scalar directions, where the bounce solution describes the motion of a soliton along a complicated manifold, requiring the use of numerical tools, such as~\cite{Konstandin:2006nd, Wainwright:2011kj, Camargo-Molina:2013qva, Masoumi:2016wot,Athron:2019nbd, Sato:2019wpo,Guada:2020xnz}.
Secondly, the stationary phase approximation used to derive the bounce action holds well for weakly-coupled theories, including radiative corrections~\cite{Langer:1969bc, Weinberg:1992ds, Buchmuller:1992rs, Gleiser:1993hf,Alford:1993br}, which assumes the existence of a hierarchy of scales, with UV and IR modes well separated in energy scales. This assumption breaks down at strong coupling, where novel methods for generalizing the saddle point treatment are necessary, as discussed in~\cite{Croon:2020cgk, Croon:2021vtc,Dupuis:2020fhh}.
Thirdly, the nucleation prefactor, often taken to be a simple $\mathcal{O}(1)$ times mass dimension 4 prefactor, can have a  non-trivial form when more carefully evaluated, and has been shown to substantially alter the nucleation rate in some cases, as its contribution may become exponential. This prefactor can be split into two independent pieces, a dynamical part, related to the inverse timescale of critical bubble growth, depending on the evolution of fluctuation of the bubble radius, as well as the thermal bath~\cite{Affleck:1980ac, Linde:1981zj,Arnold:1987mh,Csernai:1992tj,Carrington:1993ng,Moore:2000jw}, and a statistical part, which depends on functional determinants of the second-order fluctuations around the critical bubble and the symmetric phase~\cite{Baacke:1993ne,Guo:2020grp,Brahm:1993bm,Surig:1997ne,Hindmarsh:2017gnf,Croon:2020cgk}, both requiring a different formalism in order to obtain a complete treatment.

{\bf 3. Bubble wall velocity.}
Different formalisms have been developed for the calculation of $v_w$, whose applicability depends on the relative strengths of the transition, which determines  whether the terminal speed will be only mildly relativistic, or on the other hand  ultrarelativistic. For not-too-fast moving walls, a standard approach is to split the distribution functions for the various particle species in the plasma into an equilibrium part, plus a perturbation due to the interaction between the wall
and the particles.
Recently, progress has been made in characterizing the importance of the equilibrium part of the distribution function, where variation of the plasma temperature, which is a function of the position relative to the wall and $v_w$, plays a role. These variations are tied to hydrodynamic effects in the plasma, which can induce a backreaction force on the wall~\cite{Konstandin:2010dm,BarrosoMancha:2020fay,Balaji:2020yrx,Ai:2021kak,Cline:2021iff}.
For ultrarelativistic bubble walls, with a Lorentz factor $\gamma(v_w)=1/\sqrt{1-v_w^2}\gtrsim10$, equilibration cannot be maintained across the bubble wall. Nevertheless, one can assume that all the particles ahead of the advancing bubble, featuring equilibrium distributions, are absorbed by the new phase, without any reflections. The absorbed particles can then exchange momentum with the wall, which gives rise to friction. The leading effect is caused by the variation in the particles' masses across the wall due to the changing scalar condensate \cite{Bodeker:2009qy}. This gives rise to a friction force that remains independent of $v_w$, and thus cannot in general prevent a runaway behaviour towards $v_w=1$. Additionally, the particles can also emit radiation, mainly in the form of gauge bosons, which leads to a $v_w$-dependent friction effect that grows with $v_w$ and thus can prevent runaways. Single-particle emissions yield a force proportional to $\gamma(v_w)m T^3$ \cite{Bodeker:2017cim}, 
where $m$ is the mass of emitted gauge bosons inside the bubble, while a resummation of multi-particle emissions leads to an enhanced force proportional to $\gamma(v_w)^2 T^4$~\cite{Hoeche:2020rsg}, or $\gamma(v_w) m T^3$ times a log~\cite{Gouttenoire:2021kjv}.
Possible open issues include the difference between these two results and the lack of mass dependence~\cite{Azatov:2020ufh} of the force in \cite{Hoeche:2020rsg}, and the impact of radiated bosons that are reflected~\cite{Gouttenoire:2021kjv}.
Nevertheless, independent of the specific form of the friction term, the efficiency factor for the bubble wall motion can be
calculated in general~\cite{Cai:2020djd}.

With the above transition parameters $T_{\ast}$, $v_w$,  $\alpha$ and $\beta$ determined, one can go on to calculate GWs.
For a weak thermal phase transition, the dominant contribution is due to sound waves, with the GW spectrum obtained from large scale numerical simulations~\cite{Hindmarsh:2015qta,Hindmarsh:2017gnf}.
The sound shell model~\cite{Hindmarsh:2016lnk,Hindmarsh:2019phv} has been proposed to understand these numerical results, with a generalization to the expanding Universe given in~\cite{Guo:2020grp}. In this model, the 
velocity field of the perturbed plasma is modelled by a linear superposition of individual
disturbance from each bubble which in turn 
can be solved from a hydrodynamic analysis~\cite{Espinosa:2010hh}.
The resulting spectrum agrees reasonably well with that from large scale numerical calculations~\cite{Hindmarsh:2017gnf}.
Aside from the spectral shape, which does not agree perfectly with numerical result, the amplitude is
also different due to several reasons.
Firstly, the amplitude depends on the root mean square fluid velocity $\bar{U}_f$, calculable from
the hydrodynamic analysis. However,
$\bar{U}_f$ calculated this way gives an overestimation as observed in numerical simulations for strong phase transition
where $\alpha \sim 1$ with small $v_w$~\cite{Cutting:2019zws}. 
This reduction is more pronounced for increasingly smaller $v_w$ for fixed $\alpha$, due presumably to the formation of 
bubble droplets ahead of the wall which then slows down the wall. 
Secondly, the original widely used GW spectrum (see, e.g., \cite{Caprini:2015zlo}) actually enforces an infinite lifetime, $\tau_{\text{sw}}$, of sound waves, as found out in \cite{Guo:2020grp,Ellis:2018mja}. For a finite $\tau_{\text{sw}}$, an additional multiplication factor needs to be added to account for 
the increasingly reduced GW production due to the increasingly diluted 
energy density as the universe expands. This
factor depends on the expansion rate of the universe 
during the transition, and for radiation
domination it is $(1-1/\sqrt{1+2\,\tau_{\text{sw}}H})$ with $H$ the Hubble rate at $T_{\ast}$~\cite{Guo:2020grp}, which approaches the asymptotic value 
of 1 as $\tau_{\text{sw}}\rightarrow \infty$ recovering the old result, and reduces
to $\tau_{\text{sw}} H$~\cite{Ellis:2018mja,Ellis:2020awk} for short transitions.
There remains the question of what exactly 
the value of $\tau_{\text{sw}}$ is. It is usually chosen to be the time scale corresponding to the onset of turbulence \cite{Pen:2015qta,Hindmarsh:2017gnf}, which  needs to be improved based on insights gained from numerical simulations and analytical studies.
Besides, there are attempts~\cite{Giese:2020znk,Wang:2020nzm}
of going beyond the bag model~\cite{Giese:2020znk}.

We now return to the less-thermal transitions, where the vacuum energy released in phase transitions can far exceed the surrounding radiation energy (see, e.g., ~\cite{Randall:2006py,Espinosa:2008kw}).
Here the bubble expansion mode has two possibilities~\cite{Espinosa:2010hh}: strong detonation, where the wall reaches a terminal velocity due to balancing between the outward pressure and the friction, and runaway, where the wall continues to accelerate until it collides.
In determining which of the two is relevant, the friction from the thermal plasma splitting upon impinging onto the ultrarelativistic walls plays a crucial role~\cite{Bodeker:2017cim,Hoeche:2020rsg,Gouttenoire:2021kjv}.
Since this friction increases as the wall accelerates, runaway is now known to require a stronger transition than previously thought.
The main contribution to the energy budget of these transitions comes from a highly relativistic and concentrated fluid around the bubbles in strong detonations, while it comes from relativistic walls in runaway~\cite{Ellis:2019oqb}.
The GW production for the latter has long been estimated with the so-called envelope approximation, in which the walls immediately dump upon collision~\cite{Kosowsky:1992rz,Kosowsky:1992vn,Huber:2008hg,Jinno:2016vai,Jinno:2017ixd,Zhong:2021hgo,Megevand:2021llq}.
However, numerical calculations revealed that the energy accumulated on the bubble surface propagates inside other bubbles even after collision~\cite{Weir:2016tov,Cutting:2018tjt,Jinno:2019bxw,Cutting:2020nla,Lewicki:2020azd}.
To incorporate the long lifetime of these walls, a modelling now called the bulk flow model is proposed~\cite{Jinno:2017fby,Konstandin:2017sat,Megevand:2021juo}, and it is found that GWs at low frequencies are amplified, reflecting the expanding spherical structure after collision.
On the other hand, the GW production in strong detonations leaves much room for study.
While the sound shell model~\cite{Hindmarsh:2016lnk,Hindmarsh:2019phv} is expected to describe the GW production if the system turns into weak compression waves $\gamma \lesssim {\cal O} (1)$ at an early stage, it should be noted that strong concentration of the fluid may take some time to get dispersed~\cite{Jinno:2019jhi}, or 
the system may develop vortical and/or acoustic turbulence at an early stage~\cite{Caprini:2015zlo,RoperPol:2019wvy,Ellis:2020awk,Dahl:2021wyk}.

In addition, both purely hydrodynamic and magneto-hydrodynamic (MHD) turbulence are expected to source GWs \cite{Kamionkowski:1993fg}.
Past analyses have evaluated the GW production using semi-analytical modelling \cite{Kosowsky:2001xp,Dolgov:2002ra,Caprini:2006jb,Gogoberidze:2007an,Kahniashvili:2008pe,Caprini:2009yp,Niksa:2018ofa}.
The results of these analyses have been combined with the prediction of the GW signal from the acoustic phase, leading to a GW spectrum where the acoustic contribution dominates at the peak while turbulence, believed to be sub-leading, dominates at high frequencies (see e.g.~\cite{Caprini:2015zlo}). 
Simulations featuring the scalar field evolution coupled to the relativistic fluid dynamics started only recently to explore the non-linear regime, where vorticity and turbulent generation is expected to occur \cite{Cutting:2019zws}. 
Simulations of MHD turbulence carried out with the Pencil code have improved on previous analytical estimates, but have shown that the initial conditions of the turbulence onset affect the GW spectral shape around the peak region \cite{RoperPol:2019wvy,Kahniashvili:2020jgm,RoperPol:2021xnd}. 
A ready-to-use prediction of the GW signal from MHD turbulence, validated by numerical simulations but assuming fully-developed turbulence as initial condition, has been provided in \cite{RoperPol:2022iel}. 
Simulating the onset of turbulence directly from the PT dynamics, and thereby providing a thorough and reliable estimate of the GW power spectrum, remains a key challenge of the next decade.


Direct experimental searches for GWs from FOPTs have recently been carried out by several experimental collaborations. A subgroup of the LIGO-Virgo-KAGRA collaboration has performed the first search using its data from the O1, O2 and O3
observing runs \cite{Romero:2021kby}, being sensitive to FOPT at the energy scale of
$\sim $(PeV-EeV), and found no evidence for such signals, with upper limits thus placed on the FOPT parameters.
Searches have also been performed by the NANOGrav collaboration, based on its 12.5 year data set \cite{NANOGrav:2021flc}, corresponding to the QCD energy scale \cite{Witten:1984rs,Caprini:2010xv}, after the detection of a common-noise possibly due to a SGWB~\cite{NANOGrav:2020bcs}: it concludes  that a FOPT
signal would be degenerate with that from supermassive black hole binary mergers. 
A search on Parkes PTA data is also reported~\cite{Xue:2021gyq} with no positive detection and with upper limits set.
Continued searches by these efforts will give improved results in the near
future, while in a longer term, future third generation ground-based detectors such as 
Cosmic Explorer~\cite{Reitze:2019iox,Evans:2021gyd} and Einstein Telescope~\cite{Punturo:2010zz,Maggiore:2019uih} will probe much more 
weaker GW signals, and future space interferometers LISA~\cite{LISA:2017pwj,Caprini:2015zlo}, Taiji~\cite{Hu:2017mde,Ruan:2018tsw,Taiji-1} and Tianqin~\cite{TianQin:2015yph,Luo:2020bls,TianQin:2020hid} will operate in a frequency range suitable to test FOPTs at the electroweak scale. 
Once the GW signal from a FOPT is detected, recent analyses suggest that all four parameters determining its spectral shape can be reconstructed from the power spectrum, within the sound shell model \cite{Gowling:2021gcy}. 

\section{Topological Defects\label{sec:defects}}

\noindent\textbf{Motivation.} Topological defects are generically predicted in field theories with symmetry breaking \cite{Jeannerot:2003qv}  as well as superstring theories \cite{Sarangi:2002yt,Jones:2002cv}.

When a symmetry is spontaneously broken in the early Universe, the homotopy groups of the resultant vacuum manifold can be non-trivial. Consequently, topological defects of different forms can form \cite{Kibble:1976sj,Vilenkin:2000jqa}. Three types of topological defects have been shown to produce SGWB signals: domain walls \cite{Vilenkin:1981zs,Sakellariadou:1990ne,Sakellariadou:1991sd,Gleiser:1998na,Hiramatsu:2010yz,Dunsky:2021tih}, textures \cite{Fenu:2009qf} and cosmic strings \cite{Vachaspati:1984gt,Blanco-Pillado:2017oxo,Blanco-Pillado:2017rnf,Ringeval:2017eww,Vilenkin:1981bx,Hogan:1984is,Siemens:2006yp,DePies:2007bm,Olmez:2010bi,Vachaspati:2015cma}. Monopoles can also indirectly produce GWs, particularly when combined with other defects \cite{Dunsky:2021tih}.  What makes such signals particularly interesting, is that, in all cases the amplitude of the GW signal grows with the symmetry breaking scale. This makes topological defects an effective probe of high energy physics.

The type together with the detailed properties of defects which are formed depends on the underlying theory. Domain walls can in principle arise wherever there is reason to expect a discrete symmetry --- from Pecci Quinn \cite{Harigaya:2018ooc,Craig:2020bnv}, R-Parity \cite{Borah:2011qq} to neutrino masses \cite{Ouahid:2018gpg} for a non-exhaustive list. Discrete symmetries also appear naturally in the center of symmetry breaking chains --- for example D parity and matter parity \cite{Kibble:1982ae,Kibble:1982dd}. Global strings can originate from axion dark matter models where a U(1) is broken to a vacuum with a discrete symmetry \cite{Chang:2019mza}. Local strings, monopoles and textures are ubiquitous in the symmetry breaking chains that result from SO(10) breaking to the SM \cite{Dunsky:2021tih}. 
Considering all possible spontaneous symmetry breaking patterns from the GUT down to the SM gauge group, it was shown \cite{Jeannerot:2003qv} that cosmic string formation is unavoidable. The strings which form at the end of inflation have a mass which is proportional to the inflationary scale. Sometimes, a second network of strings form at a lower scale.

In the context of string theory, cosmic superstrings \cite{polchinski} can form as the result of brane interactions. While  solitonic cosmic strings are classical objects, cosmic superstrings are quantum ones, hence one expects several differences between the two  (see, e.g., \cite{Sakellariadou:2009ev} and references therein). 

\noindent\textbf{Gravitational wave signals.} 
The GW spectrum resulting from the topological defect will have different features depending upon whether the symmetry broken is local or global. For instance, GW spectrum from local textures has an infrared suppression compared to global textures \cite{Dror:2019syi}, GW signal strength from global strings features a more dramatic scaling with the symmetry breaking scale compared to the linear scaling of local strings \cite{Chang:2019mza} and local domain walls require destruction via strings in order to be viable, in contrast with global domain walls \cite{Dunsky:2021tih}. In the following we summarize the current status on GW signal from topological defects, with an emphasis on cosmic strings, which are most studied.

\noindent\textit{1. GW from local cosmic strings.}
For local strings, and particularly thin strings with no internal structure which can be described by the Nambu-Goto action, once the string network is formed, it is expected to quickly reach the scaling regime \cite{Kibble:1976sj}. The predicted GW spectrum is then contingent on two crucial quantities: the dimensionless power spectrum for a loop of a given length; and the number density of loops. Regarding the  power spectrum, either one can motivate an averaged power spectrum (where the average is over different configurations of loops of the same length) using simulation data as input \cite{Blanco-Pillado:2017oxo}, or one can assume a high frequency domination \cite{DV2,Ringeval:2017eww}. Regarding the number density,  a (rather crude) analytic estimate can be made using the `velocity dependent one scale model' that takes only the loop size at formation as an input \cite{Martins:1996jp,Martins:2000cs}. This agrees quite well with some models of simulations \cite{Blanco-Pillado:2013qja,Blanco-Pillado:2011egf}, but disagrees with others \cite{Ringeval:2005kr,Lorenz:2010sm,Auclair:2019zoz} which predict a greater fraction of energy density in smaller loops. The reasons for these differences are not yet fully understood, but may be related to the effects of gravitational backreaction.  
Besides their differences, these string models all predict a roughly constant GW spectrum over many decades of frequency, assuming standard cosmological history (see e.g.~\cite{Auclair:2019wcv}). This makes it a useful witness to any departures from a standard cosmological picture \cite{Cui:2017ufi,Cui:2018rwi, Gouttenoire:2019kij}. 
Nambu-Goto dynamics may not apply to all types of cosmic strings, in particular the case with field theory strings is unclear. Some simulation results \cite{Vincent:1996qr,Hindmarsh:2011qj} show that the field theory strings decay predominantly into particles rather than gravitational radiation, although the literature did not yet converge \cite{Matsunami:2019fss,Saurabh:2020pqe,Hindmarsh:2021mnl,Auclair:2019jip}. Such discrepancies lead to different observational predictions. Therefore it is important to further investigate this open question and better understand the difference between Nambu-Goto and field theory results. 

\noindent\textit{2. GW from global/axion cosmic strings/domain walls.}
While most of literature focused on the evolution of local cosmic strings, motivated by the close connection to axion physics, recent years have seen increasing interest in global or axion strings/topological defects and the GW signature sourced by them \cite{Sakellariadou:1991sd,Martins:2018dqg,Gorghetto:2018myk,Buschmann:2019icd,Figueroa:2020lvo,Gorghetto:2021fsn,Chang:2019mza,Chang:2021afa}. Although the GW signal from global strings is suppressed relative to local strings due to the dominant emission of Goldstone bosons, it has been shown to be detectable with upcoming experiments, and feature a logarithmic declining spectrum towards high frequency \cite{Chang:2019mza,Chang:2021afa,Gorghetto:2021fsn}.  Clarification of the GW spectrum from global strings will require further investigations in simulation studies, which so far have not converged well.

\noindent\textit{3. GW from superstrings.}
Evolution of cosmic superstring networks, is a rather involved issue, which has been addressed by numerical \cite{Sakellariadou:2004wq,PhysRevD.71.123513,Hindmarsh:2006qn,Urrestilla:2007yw,Rajantie:2007hp,Sakellariadou:2008ay}  as well as analytical \cite{Copeland:2006eh,PhysRevD.75.065024,PhysRevD.77.063521,Avgoustidis:2014rqa} approaches. Cosmic superstrings can also lead to gravitational waves (see, e.g., \cite{2006AAS...209.7413H,LIGOScientific:2017ikf}), hence GW experiments can provide a novel and powerful way to test string theory.

\noindent\textbf{Probe for non-standard cosmology.} 
The state and particle content of our Universe prior to the BBN era remains unknown as the ``primordial dark age'' \cite{Boyle:2005se,Boyle:2007zx}, despite the standard paradigm we often assume. Potential deviations from the standard cosmology scenario are well motivated and attracted increasing interest in recent years. GW background spectrum from a cosmic string network typically spans over a wide frequency range with detectable amplitude, making it a unique tool for ``cosmic archaeology" based on a time-frequency correspondence \cite{Cui:2017ufi, Cui:2018rwi}. In the following we review a few representative cases on how the GW signal from strings may be used to probe the pre-BBN cosmology.

\noindent\textit{1. Probe new equation of state of the early universe.}
In standard cosmology, the Universe undergoes a prolonged radiation dominated era from the end of inflation till the recent transition to matter domination at redshift $\sim 3000$. However, well-motivated theories suggest that the evolution of the Universe's equation of state may deviate from this paradigm, e.g. the presence of an early matter-domination or kination phase \cite{Moroi:1999zb,Nelson:2018via,Salati:2002md,Chung:2007vz, Poulin:2018dzj}. Such an alternative cosmic history can sharply modify the GW spectrum from cosmic strings via its effect on Hubble expansion rate. Specifically, an early period of kination results in a period where the GW frequency spectrum grows as $f^1$, whereas an early period of matter domination results in a spectrum that depletes in the UV, obeying a $f^{-1/3}$ power law \cite{Cui:2018rwi,Cui:2019kkd,Blasi:2020mfx}. A string network can also be ``consumed'' through the nucleation of monopoles \cite{Buchmuller:2021mbb} or domain walls if there is a small hierarchy between symmetry breaking steps, or if not they can be destroyed by a connected domain wall in a later symmetry breaking step \cite{Dunsky:2021tih}. Importantly, all of these signals can be distinguished.

\noindent\textit{2. Probe new particle species.}
While the high frequency range of the GW spectrum from strings is largely flat (corresponding to GW emission during radiation domination), it is modified by changes in the number of relativistic degrees of freedom, $g_\ast$ \cite{Cui:2017ufi, Auclair:2019wcv}, which modifies the standard Hubble expansion history, and can therefore be used as a probe of high energy degrees of freedom that are beyond the reach of terrestrial colliders or CMB observatories. 

\noindent\textit{3. Probe (pre-)inflationary universe.}
Cosmic defects generally dilute more slowly than radiation. Even if a large number of e-foldings during inflation largely washes out a pre-existing string network, it can regrow back into the horizon and replenish itself to become a non-trivial component of the late Universe energy budget. In particular, replenished strings can leave a unique SGWB spectrum that can be probed by nanoHz detectors, along with GW burst signals \cite{Cui:2019kkd}. This provides a unique example that cosmology before the end of inflation can be probed with GWs from cosmic defects.

\noindent\textbf{Probe for new particle physics.} 
As the amplitude of the GW spectrum produced by a string network grows with the symmetry breaking scale, they provide a unique way of probing high scale physics. For example, since there is typically a hierarchy between the seesaw scale and the GUT scale, it is natural to protect the seesaw scale with gauge symmetry. Symmetry breaking chains predict strings more often than not and the entire range of scales relevant to thermal leptogenesis is projected to be testable by future detectors \cite{Dror:2019syi}. More generally, GUT symmetry breaking chains more often than not allow for observable signals from some set of defects \cite{Dunsky:2021tih}.\footnote{see also Snowmass white paper \cite{Elor:2022hpa}} Searches for proton decay provide a complimentary probe of symmetry breaking chains, as chains that allow for proton decay can be chains that do not predict strings \cite{King:2020hyd,Chun:2021brv}. Furthermore, there has been a growing interest in GWs from global/axion topological defects due to the close connection to axion or axion-like (ALP) dark matter physics \cite{Chang:2019mza,Gouttenoire:2019kij,Ramberg:2020oct,Chang:2021afa,Gorghetto:2021fsn,Gelmini:2021yzu}: when PQ symmetry breaking occurs after inflation, these topological defects are an indispensable companion of axion particles. While there have been extensive studies on axion particle detection which strongly depend on whether/how axions couple to SM particles in an observable manner, the GW signature from axion strings/domain walls is universal and could be a highly complementary probe for axion physics. In addition, as mentioned earlier, the GW spectrum from cosmic strings may reveal new particle species via the effect on the Hubble expansion rate due to changes in relativistic degrees of freedom.

\noindent\textbf{Prospect for experimental detection.} 
Current limits on cosmic strings from GW signals are: $G \mu \lesssim 9.6 \times 10^{-9}$ by LIGO-Virgo \cite{LIGOScientific:2021nrg}, and $G \mu \lesssim 10^{-10}$ by pulsar timing arrays \cite{Ellis:2020ena,Blanco-Pillado:2021ygr}. 
Fig. \ref{fig:sample} shows and example of a cosmic string GW spectrum in comparison with existing and future detector sensitivities. Note that considering the expected astrophysical background and a galactic foreground, a cosmic string tension in the range of $G \mu \approx 10^{-16}- 10^{-15}$ or bigger could be detectable by LISA, with the galactic foreground affecting this limit more than the astrophysical background \cite{Boileau:2021gbr,Auclair:2019wcv}.

Future experiments covering a wide frequency range will further improve the sensitivity to GW signals from cosmic strings, including: e.g. Einstein Telescope, Cosmic Explorer, AEDGE, DECIGO, BBO, $\mu$Ares and Theia \cite{Punturo:2010zz,Yagi:2011wg,AEDGE:2019nxb,Hild:2010id,Sesana:2019vho,Theia:2017xtk}.
 An exciting possibility is that NANOGrav may have already seen a hint of cosmic strings \cite{NANOGrav:2020bcs}. The suggested hint is consistent with a shallow power law as one would expect from strings \cite{Ellis:2020ena,Blasi:2020mfx,Datta:2020bht,Chakrabortty:2020otp,Samanta:2020cdk,King:2020hyd,Garcia-Bellido:2021zgu} though to fully verify one way or another, the Hellings-Downs curve needs to be constructed \cite{Hellings:1983fr}. 

\section{Dark Matter\label{sec:dm}}

There are multiple strong and independent lines of evidences on the existence of dark matter (DM). However its identity remains as one of the greatest mysteries in the modern physics. For example, the mass of DM can range for almost 100 orders of magnitude. 
Understanding the DM identity provides invaluable information about the early Universe as well as the extension of the particle Standard Model.

Decades of effort have been devoted to searches of DM. Motivated by the gauge hierarchy problem, experimental efforts has been focused on the mass window around the electroweak energy scale, i.e. $\sim$100 GeV. 
Null results to date have led to strong constraints in this part of the parameter space and have prompted a re-examination of the possibilities of other well-motivated mass windows.

Besides conventional DM search methods, GW experiments may provide completely novel opportunities to search for DM. Interestingly, it has been demonstrated that GW experiments can be used to study DM in both ultraheavy and ultralight mass regimes, for an indirect as well as a direct detection.
(For synergy, see Snowmass white paper ``New Horizons: Scalar and Vector Ultralight Dark Matter''~\cite{Antypas:2022asj},
and ``Ultraheavy particle dark matter''~\cite{Carney:2022gse}.
)

$\bullet$ {\bf Primordial black holes (PBHs):}
GW observations~\cite{Abbott:2016blz,TheLIGOScientific:2016pea,Abbott:2016nmj,Abbott:2017vtc,Abbott:2017oio,Abbott:2017gyy,LIGOScientific:2018mvr,LIGOScientific:2020stg,Abbott:2020uma,Abbott:2020khf} have revealed intriguing properties of BH mergers
and
have rekindled suggestions that PBHs may exist and constitute a fraction of the DM~\cite{Bird:2016dcv,Clesse:2016vqa,Sasaki:2016jop,2016ApJ...823L..25K,Blinnikov:2016bxu}.
Advanced LIGO and Virgo, and future ground-based GW observatories, e.g. Cosmic Explorer (CE)~\cite{Evans:2016mbw,Reitze:2019iox} and Einstein Telescope (ET)~\cite{Punturo:2010zza,Hild:2010id,Maggiore:2019uih}, will probe the origin of BHs (stellar or primordial) through different methods and observations:
	
\textit{1. Subsolar black holes mergers.}  Detecting a black hole of mass below the Chandrasekhar mass would almost unambiguously point towards a primordial origin.  Subsolar searches have been carried out in the first three runs of LIGO/Virgo~\cite{Abbott:2018oah,Authors:2019qbw,LIGOScientific:2021job,Nitz:2020bdb,Phukon:2021cus,Nitz:2021vqh,Nitz:2022ltl}, with a few candidates recently found~\cite{Phukon:2021cus}, while
CE and ET will reach the sensitivity at cosmological distances.


\textit{2.  BHs in the NS mass range, low mass gap and pair-instability mass gap.}  
Multi-messenger astronomy could probe the origin of compact objects in the possible mass-gap between NS and astrophysical BHs \cite{Abbott:2020uma,Abbott:2020khf}, eventually revealing their possibly primordial origin by detecting EM counterparts\cite{Unal:2020mts}. CE and ET could also detect their different merging phase. PBHs in this range are motivated by a boosted formation at the QCD transition~\cite{Byrnes:2018clq,Carr:2019kxo}.  
Above $60 M_\odot$, pair-instability  should prevent BHs to form  
while PBHs could explain recent observations~\cite{Carr:2019kxo,Clesse:2020ghq,DeLuca:2020sae},
though hierarchal mergers remain a more natural explanation~\cite{Gerosa:2021mno}.  Accurate spin reconstructions allow distinguishing them from secondary mergers in dense environments~\cite{Farmer:2019jed}.   CE will probe intermediate-mass black hole binaries up to $10^4 M_\odot$,
   which will reveal a possible primordial origin of the seeds of the super-massive BHs at galactic centers~\cite{Clesse:2015wea,Carr:2019kxo}.    

\textit{3. BH mergers at high redshift.}  The third generation of GW detectors like CE and ET will have an astrophysical reach $20 \lesssim z+1 \lesssim 100$, prior to the formation of stars. Any BH merger detection would therefore almost certainly point to a primordial origin~\cite{Ding:2020ykt}.  

\textit{4. Distinguishing PBH vs stellar BHs with statistical methods.}   
Bayesian statistical methods and model selection~\cite{LIGOScientific:2018jsj} applied to  the rate, mass, spin and redshift distributions will also help to distinguish PBHs from the stellar scenarios~\cite{Kocsis:2017yty,Ali-Haimoud:2017rtz,Clesse:2017bsw,Fernandez:2019kyb,DeLuca:2019buf,Carr:2019kxo,Gow:2019pok,Hall:2020daa,Jedamzik:2020omx,Jedamzik:2020ypm,Bhagwat:2020bzh,DeLuca:2020qqa,DeLuca:2020fpg,DeLuca:2020bjf,Wong:2020yig,Garcia-Bellido:2020pwq,Dolghov:2020hjk,Dolgov:2020xzo,Belotsky:2014kca,Mukherjee:2021ags,Mukherjee:2021itf}.  They can be used to set new limits on PBH models and reveal the existence of different black hole populations  
(PBH binaries with merging rates large enough to be detected may have formed by tidal capture in clusters~\cite{Bird:2016dcv,Clesse:2016vqa,Korol:2019jud,Belotsky:2018wph} and before recombination~\cite{Nakamura:1997sm,Raidal:2017mfl,Ali-Haimoud:2017rtz,Raidal:2018bbj,Young:2019gfc}).   

\textit{5.  GW backgrounds.}   If PBHs contribute to a non-negligible fraction of DM, their binaries generate a detectable GW background~\cite{Mandic:2016lcn,Clesse:2016ajp,Raidal:2017mfl,Wang:2016ana,Wang:2019kaf,Bagui:2021dqi,Braglia:2021wwa}, as well as close encounters~\cite{Garcia-Bellido:2021jlq}.  Its spectral shape depends on the PBH mass distribution and binary formation channel, with its amplitude comparable or higher than astrophysical sources.  The number of sources contributing may also help to identify a SGWB from PBHs~\cite{Braglia:2022icu}.  Other SGWBs may come from Hawking radiation~\cite{Arbey:2019mbc,Dong:2015yjs}, from the density fluctuations at the origin of PBH formation~\cite{Ananda:2006af,Baumann:2007zm,Inomata:2016rbd,Nakama:2016gzw,Gong:2017qlj,Clesse:2018ogk,Bartolo:2018evs,Bartolo:2018rku,Inomata:2018epa,Garcia-Bellido:2017aan,Unal:2018yaa,Unal:2020mts,Romero-Rodriguez:2021aws} or from their distribution~\cite{Papanikolaou:2020qtd,Papanikolaou:2021uhe}.  The density fluctuations also give rise to anisotropies and deformation in SGWBs of other cosmological origins through propagation effects \cite{Alba:2015cms, Contaldi:2016koz, Bartolo:2019oiq, Bartolo:2019zvb, Bartolo:2019yeu, Domcke:2020xmn}.

\textit{6. Continuous waves (CWs). }   Very light ($\lsim\mathcal{O}(10^{-10}M_\odot-10^{-3}M_\odot$)) PBH binaries would generate long-lived GWs during inspiraling, lasting at least $\mathcal{O}$(hours-days) and potentially up to thousands or million years.  
A method has been designed to search for these GWs \cite{Miller:2020kmv}, and those from mini-EMRI ones~\cite{Guo:2022sdd}, with constraints placed using upper limits from searching for quasi-monochromatic, persistent GWs in O3 from planetary and asteroid-mass PBHs \cite{Miller:2021knj,LIGOScientific:2022pjk}.  CE and ET could even detect such binaries in the solar system vicinity.

\textit{7.  GW bursts.}  These may be produced by hyperbolic encounters in dense halos~\cite{Garcia-Bellido:2017qal,Garcia-Bellido:2017knh}.  The signal frequency can lie in the frequency range of ground-based detectors for stellar-mass BHs, with a duration of order of milliseconds.  Finally, the absence of GW from kilonovae may point to neutron stars (NS) destroyed by sublunar PBHs~\cite{Fuller:2017uyd}.

\textit{8. Phase transition GWs.} PBHs can form during a first-order phase transition via trapping fermions in the false vacuum~\cite{Baker:2021nyl,Baker:2021sno,Kawana:2021tde,Huang:2022him,Marfatia:2021hcp}, bubble collisions~\cite{Hawking:1982ga,Jung:2021mku} or postponed vacuum decay~\cite{Liu:2021svg,Hashino:2021qoq}. In such scenarios, there could be correlated signals between PBHs (e.g. merger/evaporation/microlensing) and phase transition GWs.




$\bullet$ {\bf Dark photon DM: }
If DM is made up of ultralight bosons, it will behave as an oscillating classical wave, with dark photon (DPDM) being a good candidate. If DPDM is further charged under $U(1)_B$ gauge group, the DPDM background field will induce displacements on the GW interferometer's test masses, resulting in a time dependent variation on the arm length and thus a GW-like signal~\cite{Pierce:2018xmy}. The DPDM signal in the frequency domain is quasi-monochromatic, centered at the mass of dark photon $m_A$, with a very narrow frequency width $\Delta f/f\sim 10^{-6}$ from the velocity dispersion of DM halo in the Milky Way.
Thus, the search amounts to bump hunting in the frequency domain with Fourier analysis. 

Searches have been performed at LIGO, which is most sensitive to a DPDM mass of $m_A\sim 4 \times 10^{-13}$eV at its most sensitive frequency $\sim 100$Hz, using data from the first observation run (O1) \cite{Guo:2019ker} and the third (O3) \cite{LIGOScientific:2021odm}. The long coherence length of the
DPDM means the signal is correlated in multiple interferometers of LIGO and Virgo, which then allows cross-correlating the strain channels of data from pairs of interferometers to significantly reduce the noises ~\cite{Allen:1997ad}, 
while in the O3 search, a band sampled data method
is also used\cite{Miller:2020vsl}.

No evidence of DPDM signal has been found in O1 and O3 data and upper limits
are placed on the DPDM coupling with baryon number.
The O3 upper limit of squared coupling $\epsilon^2$ is best constrained to be
$1.2 (1.31)\times 10^{-47}$ at $5.7(4.2)\times 10^{-13}\text{eV}/c^2$ for the two
methods used, which is improved by a factor of $\sim 100$ for 
$m_A \sim (2-4)\times 10^{-13} \text{eV}/c^2$ compared with the O1 result, with most of the gain in sensitivity coming from taking
into account of the finite travel time of the light~\cite{Morisaki:2020gui}. 
The GW data have already probed the unexplored DPDM parameter space, and direct DPDM search using GW data became competitive compared with other fifth-force experiments at this particular mass region. 

$\bullet$ {\bf Dilaton:}
The dilaton is a promising ultralight dark matter candidate. It is naturally predicted in theories with extra dimensions, and it couples to the SM particles through the trace of the energy momentum tensor. If the dilaton plays the role of DM, its oscillation will lead to time-dependent variations of fundamental constants, such as the electron mass and the fine structure constant.

If the GW interferometers are embedded in the background of the dilaton, the oscillation of the dilaton DM would manifest as changes in the length and index of refraction of materials \cite{Stadnik2015a,Stadnik2015b,Stadnik2016}. Similarly to DPDM, dilatons would behave as a classically oscillating field, and would impart a quasi-monochromatic, persistent signal onto the detectors by physically changing the size of the beam splitter, resulting in different travel times for light coming from the x- and y-arms \cite{grote2019novel}. Therefore, the arm-length of the interferometers does not matter; instead, it is necessary to have high sensitivity to optical phase differences between the two
arms. In GEO 600, the squeezed light vacuum states of light allow for a large quantum noise reduction, more than LIGO/Virgo/KAGRA, meaning that GEO 600 is the most sensitive GW interferometer to dilaton DM. A search for dilaton DM was conducted using about a week of data from GEO 600, resulting in extremely competitive constraints on the couplings of dilatons to electrons and photons \cite{Vermeulen:2021epa}. The analysis was optimally sensitive to each logarithmically-spaced mass of the dilaton, since it employed LPSD to confine the frequency modulation induced by the dilaton on the detector to one frequency bin. This modulation results from the superposition of plane waves that compose a packet of dilatons that interacts with the detector \cite{pierce2019dark}.


$\bullet$ {\bf Axions:}
Axions are scalar particles that generally appear in various extensions of the SM \cite{Peccei:1977hh, Preskill:1982cy}. 
These hypothetical particles can be constrained in several ways. 
If axions play the role of DM,  constraints can be imposed using techniques sensitive to different interaction channels and appropriate mass ranges \cite{Graham:2015ouw}. Stellar energy-loss arguments can also lead to  constraints \cite{Ayala:2014pea}.  Axions with weak self-interactions could lead to black hole superradiance, and thus be constrained through black hole spin measurements \cite{PhysRevLett.126.151102,Gruzinov:2016hcq,Davoudiasl:2019nlo,Stott:2020gjj,Ng:2020ruv,Baryakhtar:2020gao},  polarimetric observations \cite{PhysRevLett.124.061102}, and GWs emitted by the superradiance cloud \cite{Arvanitaki:2016qwi,Zhu:2020tht,Brito:2017wnc,Tsukada:2018mbp,Palomba:2019vxe}. 

Light scalar fields can be sourced by neutron stars due to their coupling to nuclear matter, and 
affect the dynamics of binary neutron star coalescence leaving potentially detectable fingerprints in the inspiral waveform.
Calculating  the first post-Newtonian corrections to the orbital dynamics, radiated power, and gravitational waveform for binary neutron star mergers in the presence of an axion field,  it was shown \cite{PhysRevD.99.063013} that  Advanced LIGO at designed sensitivity can potentially exclude axions with mass $m_{\rm a}\lesssim 10^{-11} $ eV and decay constant $f_{\rm a}\sim (10^{14}-10^{17})$ GeV.
Analyzing the GWs from the binary neutron star inspiral GW170817 allowed  to impose constraints on axions with masses below $10^{-11}$ eV by excluding the ones with decay constants ranging from $1.6 \times 10^{16}$ GeV to $10^{18}$ GeV at a 3$\sigma$ confidence level \cite{Zhang:2021mks}.
This parameter space, excluded from neutron star inspirals, has not been probed by other existing experiments.

\section{Complementarity between Collider and GW Observations\label{sec:gwcollider}}

As discussed in previous sections, GW signals from the early Universe offer a new probe of physics beyond the SM. In particular, a phase transition in the early Universe in the $100$ GeV - $100$ TeV energy range will lead to a GW signal with a peak frequency in the mHz - Hz range, potentially accessible at future GW observatories such as LISA, {Taiji, Tianqin}. This range will also be scrutinized by the LHC (including its High Luminosity upgrade) in combination with proposed future high-energy colliders like ILC, CLIC, {CEPC}, FCC, SppC or a multi-TeV muon smasher.    
%
%
GW observatories and high-energy colliders are then highly complementary in the search for physics beyond the SM: 
the discovery of a GW signal from a phase transition in the early Universe could then guide searches for new physics at future colliders; conversely, new physics discovered at colliders could provide hints of early Universe phase transitions producing GW signals. We here discuss this complementarity, highlighting that the respective sensitivities may be very different depending on the specific incarnation of new physics, and each of the two approaches could {in principle} cover the ``blind spots" of the other. 
 
We first focus on the {possibility of an} electroweak phase transition (EWPT) as a prime example leading to potential signatures both at colliders and next-generation GW detectors. Then, we also discuss other phase transitions in the early Universe for which such complementarity could be very important.

\subsection{Electroweak Phase Transition}

    Unravelling the physics behind electroweak symmetry breaking (EWSB) is central to particle physics. It will be a leading aspect of the upcoming LHC and HL-LHC physics runs, as well as a main physics driver for future high-energy colliders. There are compelling theoretical arguments to expect new physics coupled to the SM Higgs and not far from the TeV energy scale, e.g.~in order to address the origin of the electroweak scale. Among the many questions surrounding EWSB, the {thermal history of EWSB} 
    is of particular interest. In the SM {with a 125 GeV Higgs boson}, {the EWSB transition occurs via} a smooth cross-over {rather than a {\it bona fide} phase transition}~\cite{Kajantie:1996mn}. 
    {This transition takes place at a temperature $T_\mathrm{EW}\sim 140$ GeV.}
    {Importantly}, new physics coupled to the Higgs could alter the nature of the {EWSB transition}, possibly making it {a first order  EWPT}. The existence of such a transition is a necessary ingredient for electroweak baryogenesis (see Ref.~\cite{Morrissey:2012db} amd references therein as well as Snowmass white papers \cite{Barrow:2022gsu,Asadi:2022njl}) and could provide a source for observable gravitational radiation.
    {To significantly alter the SM thermal history,}
    the new physics {mass scale} cannot lie too far above {$T_\mathrm{EW}$, nor can its interactions with the SM Higgs boson be too feeble\cite{Ramsey-Musolf:2019lsf}. Thus, }
    it will be generally possible to measure its effects at the LHC or future high-energy colliders. 

    {Collider probes rely on two classes of signatures.} On the one hand, it should be possible to directly produce and study the new particles at {some of} these facilities, given that their reach in energy spans one/two orders of magnitude beyond the electroweak scale. 
    On the other hand, new physics coupled to the Higgs tends to lead to deviations $\delta_i$ of the Higgs couplings {or other properties} from their SM values, including:
    \begin{itemize}
    \item {The Higgs trilinear self-coupling, $\lambda_{3}$.}
    Although the HL-LHC will only be mildly sensitive to this coupling ($\delta_{\lambda_3} \sim 50 \% $), future colliders could significantly improve on its measurement. In particular, 100 TeV hadron colliders (e.g.~FCC-hh or SppC) and TeV scale lepton colliders could reach a sensitivity $\delta_{\lambda_3} \sim 5 - 10 \%$. 
    \item {The Higgs-$Z$ boson coupling~\cite{Craig:2013xia,Huang:2016cjm}, which could be measured with exquisite precision (down to $\delta_{Zh} \sim 0.1 \%$) in future Higgs factories like ILC, CEPC or FCC-ee.}
    \item {Higgs boson signal strengths, which depend on the product of the Higgs production cross section and decay branching ratios. Mixing between the Higgs boson and a new neutral scalar that catalyzes a first order EWPT may lead to deviations accessible at future Higgs factories~\cite{Profumo:2014opa}}.
    \item {The Higgs-to-diphoton decay rate. If a new neutral scalar is part of an electroweak multiplet, its charged partners will contribute to this loop-induced decay, with a magnitude governed by the scalar mass and the same Higgs portal coupling responsible for a first order EWPT. Order 1-10\% deviations from the SM prediction are possible, yielding potentially observable signatures at next generation colliders. }
    \item {Possible new or \lq\lq exotic\ \rq\rq Higgs decay modes into new light particles responsible for a first order EWPT.}
    
    \end{itemize}
    
    {For a generic assessment of the discovery reach for direct and indirect signals associated with a first order EWPT -- along with an extensive set of references to model-specific studies -- see Ref.~\cite{Ramsey-Musolf:2019lsf}.}

    The two types of collider probes of new physics {that may catalyze a first order EWPT}, direct production of the new particle states {and precision measurements of Higgs properties,}
    are complementary to each other and to GW probes of the {EWSB thermal history}. 
    In the following, we discuss several concrete examples of such interplay, which illustrate the reach when combining collider searches and GW observations to probe the properties of {a possible first order
    EWPT}. 
    
    
    \noindent {\it{Singlet-driven EWPT scenarios}}. {The interactions of a SM gauge singlet scalar with the Higgs open up significant possibilities for a first order EWPT. A singlet may be either real (the \lq\lq xSM\,\rq\rq) or complex (the \lq\lq cxSM\,\rq\rq ) and involve adding to the SM one or two new degrees of freedom, respectively. 
We focus on the EWPT in the xSM (see, {\it e.g.}, \cite{Barger:2007im,Profumo:2007wc,Espinosa:2011ax,Profumo:2014opa,Huang:2016cjm}).}
    In the absence of a $\mathbb{Z}_2$-symmetry for the singlet scalar field 
$S$, the Higgs and the singlet will generally mix. {On general grounds, one expects $\vert\sin\theta\vert \gtrsim 0.01$ when a first order EWPT is sufficiently strong as to accommodate electroweak baryogenesis \cite{Ramsey-Musolf:2019lsf}. }
The presence of the singlet, both via the mixing angle $\theta$ and via its contribution to the Higgs two-point function at loop level, leads to a universal suppression of Higgs couplings to gauge bosons and fermions w.r.t. their SM values. {Precision studies of Higgs boson properties provide multiple avenues for observing these effects. For example, }
it has been shown in~\cite{Huang:2016cjm,Chen:2017qcz} (see also~\cite{Curtin:2014jma}) that the resulting modification of the Higgs coupling to the $Z$ boson would allow one to probe a large fraction of the parameter space region yielding a strongly first-order EWPT at FCC-ee, CEPC or ILC-500.  {Measurements of the Higgs boson signal strengths at the LHC or future Higgs factories could provide a similarly powerful probe, as shown in Ref.~\cite{Profumo:2014opa}.} 
The Higgs self-coupling $\lambda_3$ could be measured at a future 100 TeV hadron collider or a multi-TeV lepton collider (e.g. CLIC or a muon smasher) with $10\%$ precision {or better}, which yields a comparable constraint on the singlet parameter space in the small-mixing limit $\rm{sin}\,\theta \ll 1$~\cite{Profumo:2014opa,Chen:2017qcz}. 

{On the other hand, it is possible that the degree of singlet-Higgs mixing needed for a first order EWPT may not be entirely accessible with future precision Higgs studies. In this case, direct production via the \lq\lq resonant di-Higgs\rq\rq\, process (for $m_s > 2\, m_h$) provides a complementary approach. It was shown in Ref.~\cite{Kotwal:2016tex} that searches for this process in the channels $p p \to s\to hh\to b\bar b \gamma\gamma$ and $p p \to s\to hh\to 4 \tau$ at a 100 TeV hadron collider could cover the entire first order EWPT parameter space, including portions not accessible through precision Higgs studies. (See also Refs.~\cite{Huang:2017jws,Li:2019tfd} for resonant di-Higgs probes of the EWPT in the xSM with the $bbWW$ and $4b$ channels at the HL-LHC.) Additional direct production possibilities include vector boson fusion (VBF) production of the singlet at 3 TeV CLIC~\cite{No:2018fev,Buttazzo:2018qqp} or a multi-TeV muon collider~\cite{Buttazzo:2018qqp,Liu:2021jyc} with its subsequent decay to a Higgs boson pair ($s\to hh\to b\bar b b\bar b$) or $Z$ boson pair ($s\to ZZ\to\ell^+\ell^-\ell^+\ell^-$) (with higher collision energy setups giving higher reach). }

\begin{figure}[!t]
\begin{center}
\includegraphics[width=0.8\hsize]{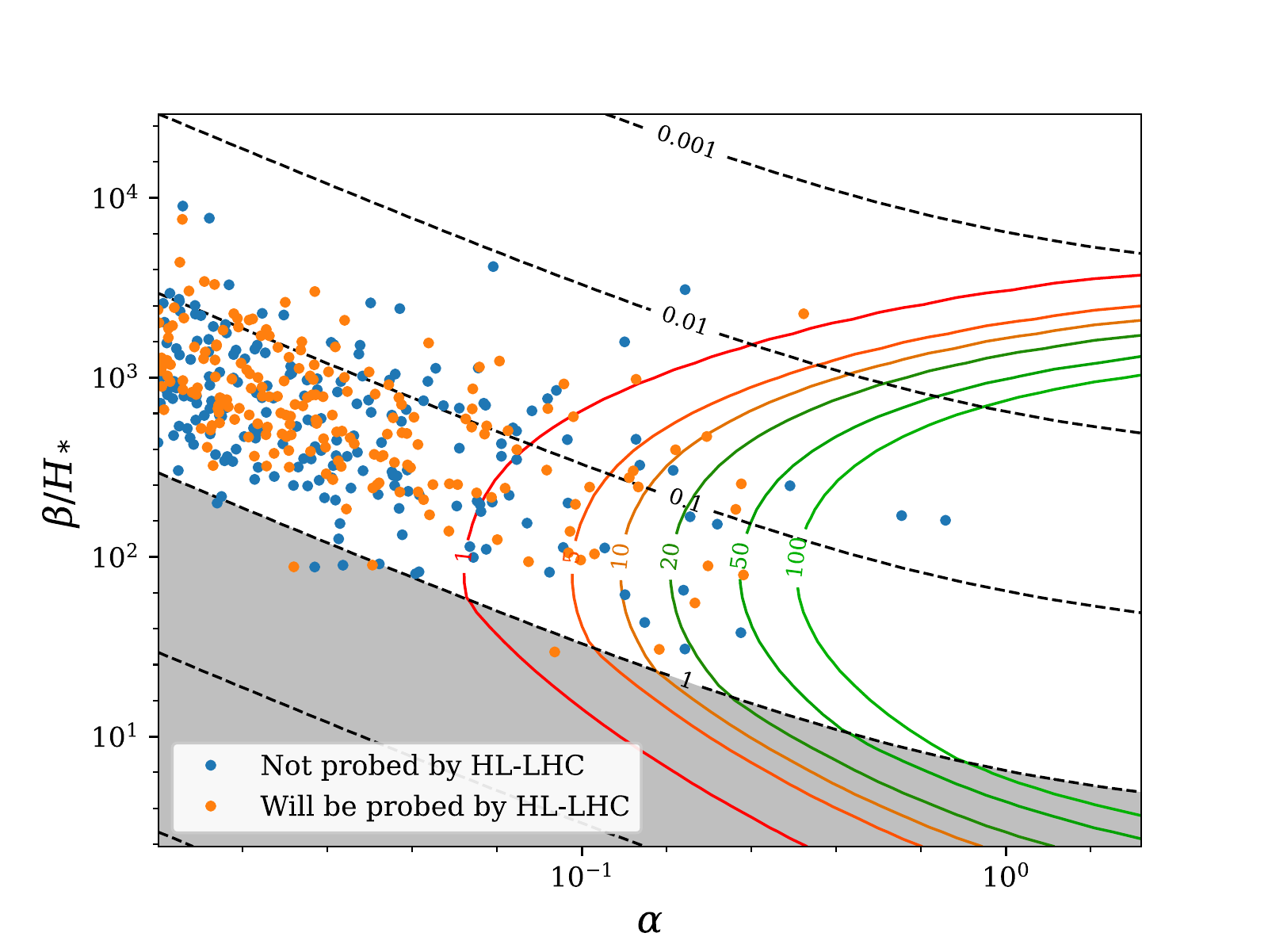}
\end{center}
\caption{
EWPT strength $\alpha$ versus inverse duration (in Hubble units) $\beta/H^{*}$ 
for xSM benchmark scenarios. The orange benchmarks feature a singlet mixing 
$\left\vert\rm{sin}\,\theta \right\vert \gtrsim 0.1$, thus within reach of the HL-LHC, 
while the HL-LHC will not be able to probe the blue points (some of which are within reach of LISA). 
The red-orange-green curves correspond to the LISA sensitivity with a certain signal-to-noise 
ratio (indicated in the figure). The black dashed lines correspond to constant values of 
$(\tau_{\text{sw}} H)^{-1}$ (see section \ref{section_PT}), with
$\tau_{\text{sw}} H < 1$ for the grey region. Figure adapted from~\cite{Caprini:2019egz} 
using \textsc{PTPlot}~\cite{PTPLOT}. 
}
\label{fig:LHC_LISA_complementarity}
\end{figure}

Conversely, for small singlet scalar masses $m_s < m_h/2$, exotic Higgs decays $h \to s s$ will also allow to probe the corresponding first-order EWPT region~\cite{Carena:2021onl,Kozaczuk:2019pet} at the HL-LHC (in the $b\bar b \tau \tau$ final state) and future lepton colliders (in the $4 b$ final state). When combined with the projected sensitivity to the EWPT via GWs of the future LISA detector, (almost) the entire parameter space yielding detectable GW signals would be probed by future multi-TeV lepton colliders~\cite{Liu:2021jyc} (and also by 100 TeV hadron colliders). Thus, if a stochastic GW signal from a phase transition were to be detected by LISA, these future collider facilities would provide a key cross-check to identify the underlying new physics. {The complementarity between GW probes with the future LISA detector and new physics searches at colliders (in this case the HL-LHC) is shown explicitly for the xSM in Fig.~\ref{fig:LHC_LISA_complementarity}, in the plane of EWPT strength $\alpha$ and inverse duration (in Hubble units) $\beta/H^{*}$ (see section \ref{section_PT}), using \textsc{PTPlot}~\cite{PTPLOT}.}

    For the $\mathbb{Z}_2$-symmetric singlet extension of the SM \cite{Profumo:2007wc,Curtin:2014jma}
    the direct signatures\footnote{Indirect signatures of the singlet through the modifications of Higgs couplings would be similar to the non-$\mathbb{Z}_2$-symmetric case discussed above in the limit of vanishing Higgs-singlet mixing.} at colliders differ from the ones discussed above: the singlet field does not mix with the Higgs, has to be produced in pairs and does not decay to SM particles, escaping the detector as missing transverse energy $E_T^{\rm{miss}}$. 
    For $m_s > m_h/2$, the sensitivity of the HL-LHC (in the VBF process $p p \to 2j + s s$ via an off-shell Higgs) will be very limited, and the parameter space yielding a first-order EWPT will only be accessible at a future 100 TeV hadron collider~\cite{Curtin:2014jma,Craig:2014lda} or multi-TeV lepton colliders~\cite{Buttazzo:2018qqp,Chacko:2013lna,Ruhdorfer:2019utl} (also possibly at a high-energy $\gamma\gamma$ collider based on such lepton colliders~\cite{Garcia-Abenza:2020xkk}). A space-based GW observatory like LISA would then have the first chance to probe the parameter space with a first-order EWPT in the $\mathbb{Z}_2$-symmetric scenario (as discussed in section 6.1 of~\cite{Caprini:2019egz}). 
    
    

    For other (non-singlet) extensions of the SM yielding a strong first-order EWPT, e.g. with new scalar electroweak multiplets, the non-singlet nature of the new fields helps making them more accessible at high-energy colliders. This strengthens the interplay between LHC studies and the generation of GWs from the EWPT. {Important cases of recent interest include:}
    
    \noindent{\it{Two-Higgs-doublet models (2HDM)}}. {In this scenario,} a first-order EWPT favours a sizable mass splitting among the new states $A_0$ and $H_0$ from the second Higgs doublet, and LHC searches for $p p \to A_0 \to H_0\, Z$ yield important constraints on the corresponding EWPT parameter space~\cite{Dorsch:2014qja,Dorsch:2016tab}. 
    The HL-LHC will completely probe the first-order EWPT parameter space in the 2HDM of Type-II (see e.g.~section 9.4 of~\cite{Cepeda:2019klc}), while for Type-I, LISA will be able to explore parameter regions beyond the LHC. 
    
    \noindent {\it{Extension of the SM by a (real) scalar $SU(2)_L$ triplet}.} {This scenario\cite{Patel:2012pi,Blinov:2015sna,Niemi:2018asa,Chala:2018opy}, which entails adding three new degrees of freedom to the SM,} allows for a very strong first-order EWPT through a two-step process~\cite{Patel:2012pi} within the reach of LISA. It also predicts various distinct collider signatures, including the modification of the $h \to \gamma\gamma$ branching fraction and disappearing track signatures~\cite{Chiang:2020rcv} associated with the compressed triplet scalar spectrum. 
    
\begin{figure}[!t]
\begin{center}
\includegraphics[width=1.0\hsize]{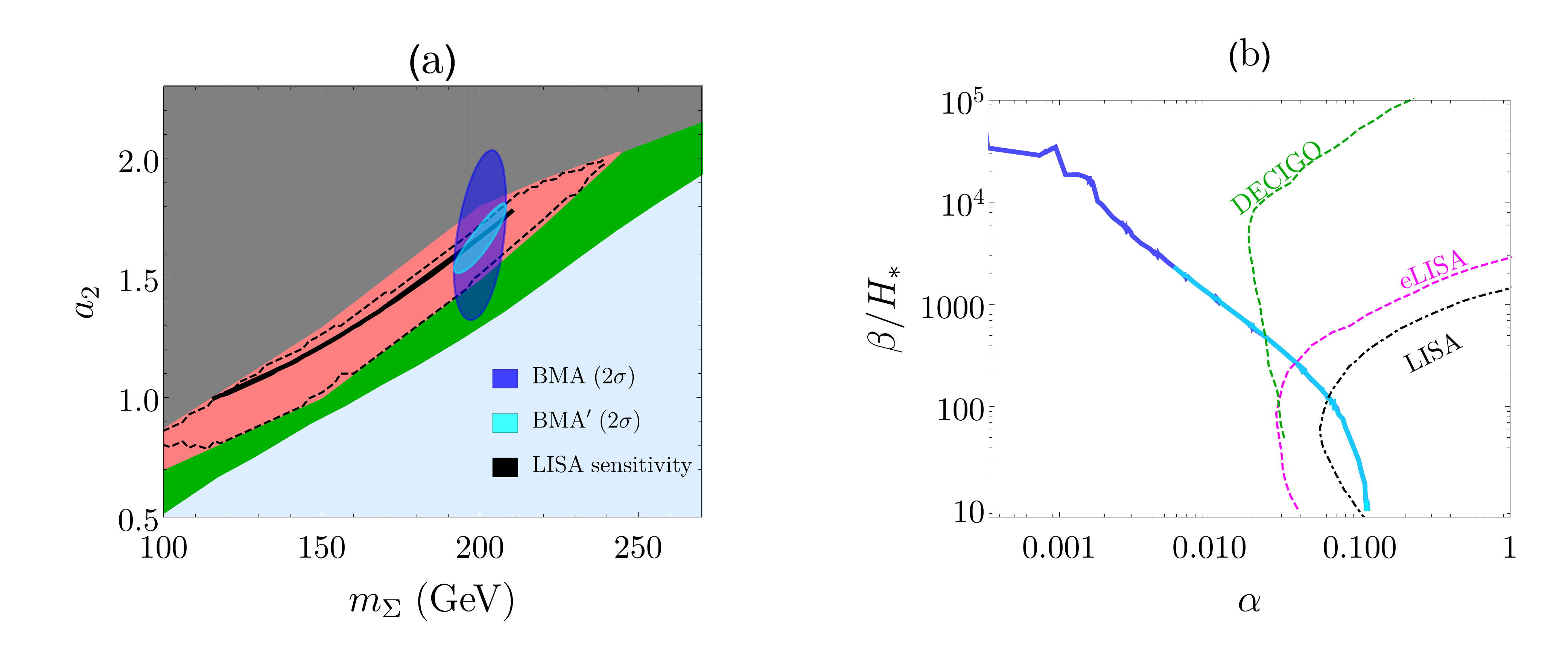}
\end{center}
\caption{Real triplet extension of the SM.  Panel (a) gives the phase diagram in terms of the triplet mass $m_\Sigma$ and Higgs portal coupling $a_2$. The light blue, green, red, and grey areas correspond to singlet step crossover transition, single step first order transition, two step thermal history, and unstable electroweak minimum, respectively. The interior of the black dashed contour corresponds to an EWPT that would complete. The thin black band is the allowed region for a hypothetical LISA observation. The dark (light) ellipses give prospective collider allowed regions for scenarios BMA (BMA'): determination of the triplet mass and Higgs diphoton decay rate (adds a measurement of the neutral triplet decay to two $Z$ bosons). The blue bands in panel (b) show projection of the hypothetical collider allowed parameter space into the plane of GW-relevant inputs. Figures adapted from Ref.~\cite{TripletGW2022}.
}
\label{fig:GW-Collider-triplet}
\end{figure}
    
    {Recent work on the triplet model that draws on the use of effective field theory and lattice simulations, illustrates how the combination of GW and collider observations could test the model and identify the values of the relevant parameters\cite{TripletGW2022}. Illustrative results are shown in Fig. \ref{fig:GW-Collider-triplet}. Panel (a) shows the phase diagram (see caption for details), with the thin black band giving the portion region that a hypothetical GW signal would identify. The dark and light blue ellipses correspond to the results of prospective future collider observations. An overlap between the collider and GW-allowed regions could enable a precise determination of the model parameters in a way that is unlikely to occur with either GW or Collider measurements alone. Panel (b) shows how a hypothetical collider-allowed region in the plane of GW input parameters. Some portions of that parameter space would lie within the range of currently envisioned GW probes, while a complete coverage would require development of more sensitive probes.  }


Finally, some scenarios envision the dark matter actually plays a role in modifying the nature of the EWPT. An example in this class is the so called lepton portal dark matter model, with Majorana fermionic dark matter candidate~\cite{Bai:2014osa} which has negligible direct and indirect signals. Therefore, the only hope to probe such a WIMP scenario is the collider searches. The scalar portal interactions between the charged mediator and the SM Higgs are first considered in Ref.~\cite{Liu:2021mhn}, which shows that the EWPT can be modified to a first-order one, providing an extra detecting channel, i.e. the first order EWPT GWs. It is shown that the precision measurement at the future CEPC on Higgs exotic decay and Higgs coupling to lepton can be complementary to the GW signals in probing the scalar and lepton portal couplings.

\subsection{Theoretical Robustness}

When exploring the EWPT sensitivity of future GW and collider probes, it is important to assess the reliability of the theoretical computations. To date, the bulk of such studies have relied on the use of perturbation theory to analyze the EWPT thermodynamics and bubble nucleation dynamics (in the case of a first order EWPT). However, it has long been known, if not widely appreciated in recent times, that exclusive reliance on perturbation theory can yield quantitatively and qualitatively misleading results. The most significant impediment arises from bosonic contributions to thermal loops. While non-perturbative lattice computations are free from this difficulty, exclusive reliance on the lattice is not practical for a wide survey of BSM scenarios and the relevant parameter space therein. A reasonable middle ground is to pair the lattice with state-of-the-art perturbative computations, using the former to \lq\lq benchmark\rq\rq\, the latter. The latter now relies on the use of the dimensionally-reduced, three-dimensional effective field theory (DR 3DEFT). For recent applications to the singlet, 2HDM, and real triplet models, see Refs.~\cite{Gould:2019qek,Niemi:2021qvp,Kainulainen:2019kyp,Niemi:2018asa,Niemi:2020hto}.

A recent application of this benchmarking approach to the GW-collider interplay is given in Ref.~\cite{Gould:2019qek}. In that study, it was observed that if a new scalar is sufficiently heavy to be integrated out of the DR 3DEFT, yielding a high-temperature effective theory that is SM like, then the resulting GW signal is unlikely to be accessible with LISA. Only for a sufficiently light or dynamical new scalar could one expect a signal in the LISA detector. On the other hand, future collider searches could still probe the first order EWPT-viable models that would be inaccessible to LISA. Looking to the future, the use of lattice-EFT methods to investigate the prospects for next generation GW probes is a clear theoretical forefront.


\subsection{Other Phase Transitions}

Other phase transitions close to the weak scale can also leave both collider and GW signal. One of the well motivated scenarios is Supersymmetry, which
could be probed with GWs, complementing the existing efforts in collider physics.

First, the presence of light supersymmetric particles (specifically the stops) coupled to the SM Higgs could affect the properties of the EW symmetry breaking, rendering it strongly first order \cite{Carena:1996wj,Delepine:1996vn}. However such light stop scenario is in tension with LHC data \cite{Menon:2009mz,Cohen:2012zza,Curtin:2012aa,Carena:2012np}.
Within the same strategy, one could consider non minimal SUSY extension of the SM including extra scalars which further favour a strong FOPT, without being excluded by the LHC (see e.g. \cite{Huang:2014ifa,Kozaczuk:2014kva,Huber:2015znp}). The phenomenology of such models possesses features similar to the singlet extensions of the SM which have been previously discussed.

A promising scenario for GW signatures in the MSSM has been investigated recently in \cite{Fornal:2021ovz}. In this realization the MSSM scalars have a non-standard thermal evolution at high temperatures, passing through a phase of symmetry non-restoration. The associated phase transition at high temperature can be a source of GWs, possibly detectable in future interferometers.

Futhermore, viable SUSY models should include a sector where SUSY is spontaneously broken, and SUSY breaking is tied to spontaneous R-symmetry breaking \cite{Nelson:1993nf}.
In \cite{Craig:2020jfv} the SUSY and R-symmetry phase transition in simple SUSY breaking hidden sectors has been investigated, and it has been shown under which conditions this can be strong and first order, leading to a SGWB. Constraints from gravitino cosmology set bounds on the SUSY breaking scales resulting in a GW spectrum in the frequency ranges accessible to current and future interferometers. Moreover, once the SUSY breaking mediation scheme is specified, the peak of the GW spectrum is correlated with the typical scale of the SM superpartners, and a visible GW signal would imply superpartners within reach of future colliders. 

As a generic remark, we emphasize that SUSY gauge theories typically include large scalar manifolds where phase transitions could have occurred during the evolution of the Universe, opening the possibilities for novel mechanisms to generate GW signatures and to test high energy SUSY breaking scales.

\section{Correlating GW Background with EM Observations\label{sec:gwem}}

As discussed above, a SGWB arises as an incoherent superposition of many GW sources, summed over all sky directions and both polarizations. Numerous SGWB models have been proposed, both cosmological and astrophysical, many of which are accessible to terrestrial and space-borne GW detectors~\cite{maggiore,regimbau_review}. These models often yield predictions for other cosmological observables, such as the CMB, the distribution of galaxies across the sky and redshift, and the distribution of dark matter throughout the universe. It is therefore expected that cross-correlating the spatial structure in the SGWB with spatial structures in other cosmological observables would enable new probes of the underlying physical models and of the earliest phases of the evolution of the universe.

The SGWB is typically described in terms of its energy density \cite{allenromano,romanocornish}:
\begin{equation}
 \Omega_{\rm GW}(\hat{e},  f)\equiv\frac{f}{\rho_c} \frac{\rm d^3\rho_{\rm GW}\left(f, \hat{e}\right)}{{\rm d} f{\rm d}^2 \hat{e}}
 = \frac{\bar{\Omega}_{\rm GW}}{4\pi}+\delta\Omega_{\rm GW}(\hat{e},  f) ,
\end{equation}
where $d \rho_{\rm GW}$ is the energy density of gravitational radiation stored in the frequency band
$[f, f+df]$, $\hat{e}$ is the direction on the sky, and $\rho_c$ is the critical energy density needed for a spatially flat Universe. In the second step, we have separated the isotropic and anisotropic components of the SGWB energy density. The anisotropic part can further be decomposed in spherical harmonics $\delta\Omega_{\rm GW}(\hat{e},  f) =  \sum_{lm} a_{lm}(f) Y_{lm}(\hat{e})$, from which the angular power spectrum can be computed as
$C_{l}(f) \propto \sum_m \langle a_{lm}(f) a^{*}_{lm}(f) \rangle$, under the assumption of statistical isotropy. 

While the isotropic SGWB component is expected to be larger than the possible anisotropy across the sky, there have been significant recent developments in the literature computing the levels of anisotropy for various astrophysical and cosmological SGWB models
\cite{contaldi,jenkins_cosmstr,Jenkins:2018a,Jenkins:2018b,Jenkins:2019a,Jenkins:2019b,Jenkins:2019c,Cusin:2017a,Cusin:2017b,Cusin:2018,Cusin:2018_2, Cusin:2019,Cusin:2019b,Alonso:2020,Cusin:2019c,Cusin:2018avf,CanasHerrera,ghostpaper,bartolo,bartolo2020,Adshead:2020bji,dallarmi,Bellomo:2021mer,domcke2020,ricciardone2021,braglia2022}. Some of them have also investigated the possibility of correlating the SGWB anisotropy with the anisotropy observed in electromagnetic (EM) tracers of the large scale structure, such as galaxy counts and weak lensing \cite{Cusin:2017a,Cusin:2017b,Cusin:2018,Cusin:2019,CanasHerrera,Alonso:2020,Cusin:2019c,mukherjee, Alonso:2020mva,PhysRevD.104.063518}, or the CMB \cite{ghostpaper,contaldi,bartolo,bartolo2020,dallarmi,domcke2020,ricciardone2021,braglia2022}. 
In such cases, one can also expand the EM observable (such as galaxy count distribution) in spherical harmonics (with coefficients $b_{lm}$) and define the angular cross-correlation spectrum $D_{l}(f) \propto \sum_m \langle a_{lm}(f) b^{*}_{lm}(f) \rangle$. These SGWB-EM anisotropy correlations carry unique potential to probe different aspects of high-energy physics, as we outline in the following examples.

{\bf SGWB-CMB Correlations:} While most cosmological SGWB models predict isotropic backgrounds ~\cite{grishchuk,barkana,starob,turner,peloso_parviol,seto,eastherlim,boylebuonanno}, recent studies have started to investigate anisotropy in these models. An example is the model of phase transitions (PT) in the early universe, which occurred as the universe cooled and went through symmetry-breaking transitions   \cite{witten,hogan,turnerwilczek,kosowsky,kamionkowski,apreda,caprini,binetruy,caprini2}. As bubbles of a new vacuum form and expand, collisions of bubble walls, combined with corresponding motion of the plasma and magnetohydrodynamic turbulences lead to formation of the SGWB \cite{caprini2}, as discussed in Section 3. A PT is expected to occur at the time of the electroweak symmetry breaking, at $\sim 1$ TeV scale, resulting in a potentially strong SGWB in both LISA and third-generation (3G) terrestrial detector bands~\cite{binetruy,margotpaper}. Possible PTs at higher temperatures ($\sim 10^3 - 10^6$ TeV) would also be accessible to 3G detectors~\cite{margotpaper}. The PT would have occurred at slightly different redshifts in different causally disconnected regions of the universe, giving rise to anisotropy in the SGWB. The SGWB angular structure would not be affected by interactions with the plasma (i.e. effects such as Silk damping and baryon acoustic oscillations are not relevant for GWs), resulting in a simple angular spectrum: $C_l^{\rm GW} \sim [l (l+1)]^{-1}$ \cite{ghostpaper}. Assuming the PT happened after inflation, the primordial density fluctuations that led to the CMB angular spectrum would also have been present during the PT,  imprinting a SGWB anisotropy at least as large as
the CMB anisotropy~\cite{ghostpaper}.  The degree and nature of correlations between the two backgrounds would provide valuable insight into inflation and the ''dark ages`` of cosmic history. 

While SGWB anisotropy can be generated at the time of the SGWB production, as in the above phase transition example, it is also possible for the SGWB anisotropy to be generated while GWs propagate through a non-uniform universe. This effect is common for all isotropic early-universe SGWB models: as GWs propagate through large-scale density perturbations  that are correlated with the CMB temperature and E-mode polarization, they too become correlated with the CMB \cite{ricciardone2021,contaldi,bartolo,bartolo2020,domcke2020}. Multiple examples of extensions of $\Lambda$CDM universe model have been examined, all featuring new pre-recombination physics. This  includes models with extra relativistic degrees of freedom, a massless non-minimally
coupled scalar field, and an Early Dark Energy component \cite{braglia2022}. SGWB-CMB correlations help constrain these models at various levels of significance, depending on the specific model and on the strength of the SGWB monopole.

{\bf Cosmic strings:} As discussed in Section 4, cosmic strings, either as fundamental strings or as topological defects formed during PTs in the early universe, are expected to support cusps \cite{caldwellallen,DV1,DV2,cosmstrpaper} and kinks \cite{olmez1}, which if boosted toward the Earth could result in detectable GW bursts. Integrating contributions of kinks and cusps across the entire string network results in a SGWB. Discovery of the cosmic superstring SGWB would open a unique and rare window into string theory \cite{polchinski}. The amplitude, the frequency spectrum, and the angular spectrum depend on fundamental parameters of cosmic strings (string tension, reconnection probability), and on the network dynamics model \cite{siemens,ringeval,olum}. While the isotropic (monopole) component of this SGWB may be within reach of the advanced or 3G detectors~\cite{O1cosmstr}, the anisotropy amplitudes are found to be $10^{4} - 10^{6}$ times smaller than the isotropic component, depending on the string tension and network dynamics~\cite{olmez2,jenkins_cosmstr}. This level of anisotropy may be within reach of the 3G detectors. Correlating the anisotropy of this SGWB with anisotropy in the CMB or large scale structure may reveal details about the formation and dynamics of the cosmic string network.

{\bf Primordial Black Holes (PBHs):} As discussed in Section 5, PBHs are of high interest as dark matter candidates and have been searched for using different observational approaches, including gravitational lensing, dynamical effects, and accretion effects. While constraints have been placed that disfavour PBHs as a significant fraction of the dark matter, they are far from conclusive due to the variety of assumptions involved, and consequently a window around $10 M_\odot$ is still allowed. Cross correlating the sky-map of the SGWB due to binary black hole (BBH) signals with the sky-maps of galaxy distribution or dark matter distribution could provide additional insights on the origin of black holes \cite{raccanelli1,raccanelli2,nishikawa,scelfo}. In particular, in more massive halos the typical velocities are relatively high, making it harder for two PBHs to form a binary through GW emission, since the cross section of such a process is inversely proportional to some power of the relative velocity of the progenitors. The PBH binaries are therefore more likely to form binaries in low-mass halos. On the other hand, the merger probability for stellar black holes is higher in more luminous galaxies (or more massive halos). Therefore, if the BBH SGWB anisotropy is found to be correlated with the distribution of luminous galaxies, the BBHs would be of stellar origin, otherwise they would be primordial. While mergers of PBHs would tend to trace the filaments of the large-scale structure, stellar BBH mergers would tend to trace the distribution of galaxies of high stellar mass. The clustering of well-resolved individual GW sources may provide additional constraints \cite{stiskalek2020, payne2020, bera2020}, and efforts to combine well-resolved sources with the SGWB hold promise as well \cite{callister2020}.

{\bf Outlook for GW-EM observations:} Advanced LIGO and Advanced Virgo have produced upper limit measurements of the SGWB anisotropy in the 20-500 Hz band for different frequency spectra, and for both point sources and extended source distributions on the sky \cite{O1stochdir,O2stochdir}. Similar techniques for measuring SGWB anisotropy in the 1 mHz band using LISA are being developed \cite{adamscornish}. The first attempts to correlate SGWB measurements with EM observations are also being developed (for example with the SDSS galaxy survey, resulting in upper limits on the cross-correlation \cite{yang}). Much more remains to be done in order to fully explore the science potential of the SGWB-EM correlation approach. Systematic studies are needed to understand the angular resolution of GW detector networks and to perform optimal SGWB-EM correlation measurements so as to start constraining model parameters---e.g. Bayesian techniques applied to the BBH SGWB are particularly promising \cite{smiththrane,banagiri}. Further development of theoretical models of SGWB-EM anisotropy correlation is critical to enable formulation of suitable statistical formalisms to compare these models to the data. Finally, the study of the astrophysical and cosmological components of the SGWB and their correlations with different EM observations will be further deepened by the upcoming, more sensitive data coming from gravitational wave detectors (LIGO, Virgo, Kagra, Einstein Telescope, Cosmic Explorer, LISA), galaxy and weak lensing surveys (EUCLID, SPHEREx, DESI, SKA, and others), and CMB measurements.

\section{Conclusions\label{sec:conclusion}}

This white paper highlights the strong scientific potential in using GW observations to probe fundamental particle physics and the physics of the early universe. Processes that took place in the Universe within one minute after the Big Bang are often associated with high energies that cannot be reproduced in laboratories, making GW observations unique opportunities to probe the new physics at such energies. In some cases, combining GW observations with accelerator-based experiments or with cosmological observations in the electromagnetic spectrum allows even more powerful probes of the new physics. 

The standard inflationary paradigm results in a scale-invariant GW spectrum. A novel coupling between the inflaton and gauge fields could result in a strongly blue-tilted spectrum. At the end of inflation, a variety of mechanisms for transferring energy from the inflaton to other particles, including reheating and preheating phases, could result in a boost of the GW spectrum at relatively high frequencies. The presence of additional phases in the Universe, especially if characterized by stiff equations of state, could also result in a significant blue GW spectrum observable by future GW detectors. Alternative cosmologies, such as pre-Big-Bang and ekpyrotic models, could also leave observable blue GW spectra, hence providing new windows into the origins of the universe.

As the Universe cools, multiple symmetries are expected to be broken at different energy scales, resulting in phase transitions in the early universe. The electroweak phase transition is of particular interest, but others are also possible, including QCD, supersymmetry, and others. If they are first-order, these phase transitions could result in GW production with a spectrum typically peaked around the frequency corresponding to the energy scale of the phase transition. GW production mechanisms include collisions of bubble walls, sound waves in the plasma, and magnetohydrodynamic turbulence. In the case of electroweak phase transitions, existence of new physics slightly above the electroweak scale could cause the transition to be first-order. Such new physics would be within reach of future collider experiments, including the ILC, FCC, CEPC, and others, hence raising the distinct possibility of combining collider experiments with GW observations to probe the physics of electroweak symmetry breaking. Similar collider-GW complementarity could also be used to study other symmetries and corresponding phase transitions, with supersymmetry as a notable example.

Furthermore, phase transitions in the early universe could result in topological defects, such as strings or domain walls. Among these, cosmic strings have received much attention as possible GW sources: the dynamics in cosmic string loops is expected to produce a broad GW spectrum, spanning decades in frequency, with the amplitude strongly dependent on the string tension. More recently, axion strings and topological defects have been studied as sources of GWs, likely with a spectrum with logarithmic decline at high frequencies. Cosmic superstrings are also a possible GW source, turning GW experiments into novel ways to test string theory.

Dark matter could result in GW production with a broad variety of morphologies. Dark matter in the form of primordial black holes could be detected in individual binary black hole merger events, for example if involving subsolar black holes or if taking place at high redshift ($>20$), or by observing the SGWB due to binary black holes whose spectrum would depend on the fraction of black holes that are of primordial origin. Dark matter in the form of dark photons would induce quasi-monochromatic displacements in the GW detector test masses, at the frequency set by the dark photon mass, and could be searched for using Fourier techniques. Dark matter in the form of a dilaton would cause changes in the length and index of refraction in a GW detector's mirrors, hence inducing phase differences in the detector's two arms. Dark matter in the form of axions could generate GW signatures through the black hole superradiance process or by modifying the inspiral signal in neutron star binary mergers. 

Finally, we note that cross-correlations of the anisotropy in GW energy density and the anisotropy in electromagnetic tracers of the structure in the universe (cosmic microwave and infrared backgrounds), weak lensing, galaxy counts) could also serve as powerful probes of the early universe physics. As an example, if a phase transition happened after inflation, the primordial density fluctuations that led to the CMB angular spectrum would also have been present during the phase transition, imprinting anisotropy in the SGWB at least as large as (and correlated with) the CMB anisotropy. Other applications include cosmic string probes and probes of dark matter in the form of primordial black holes. 

This tremendous breadth of fundamental particle physics and cosmology phenomena will be accessible to future GW observations, including terrestrial and space-borne detectors, pulsar timing observations, and experiments targeting the B-mode CMB polarization. Realizing this scientific potential requires not only development and completion of the next generation of GW and collider detectors, but also theoretical developments that would define effective probes of the phenomena using the upcoming data. 

\bmhead{Acknowledgments}
RC is supported in part by U.S. Department of Energy Award No. DE-SC0010386.
YC is supported in part by the U.S. Department of Energy under award number DE-SC0008541. 
HG, FY and YZ are supported by U.S. Department of Energy under Award No. DESC0009959. 
VM is supported by NSF grant PHY2110238. VM and CS are also supported by NSF grant PHY-2011675.
AM and AS are supported by the SRP High-Energy Physics and the Research Council of the VUB, AM is also supported by the EOS - be.h project n.30820817, and AS by the FWO project G006119N.
JMN is supported by Ram\'on y Cajal Fellowship contract RYC-2017-22986, and by grant PGC2018-096646-A-I00 from the Spanish Proyectos de I+D de Generaci\'on de Conocimiento. 
MJRM is supported in part under U.S. Department of Energy contract DE-SC0011095, and was also supported in part under 
National Natural Science Foundation of China grant No. 19Z103010239. 
MS is supported in part by the Science and Technology Facility Council (STFC), United Kingdom, under the research grant ST/P000258/1.
KS is supported in part by the U.S. Department of Energy under award number DE-SC0009956.
LTW is supported by the U.S. Department of Energy grant DE-SC0013642.
GW is supported by World Premier International Research Center Initiative (WPI), MEXT, Japan.
HA is supported in part by the National Key R\&D Program of China under Grant No. 2021YFC2203100 and 2017YFA0402204, the National Natural Science Foundation under Grant No. 11975134, and the Tsinghua University Initiative Scientific Research Program.
LB is supported in part by the National Key Research and Development Program of China Grant No. 2021YFC2203004, the National Natural Science Foundation of China under the grants Nos.12075041, 12047564, and the Fundamental Research Funds for the Central Universities of China (No. 2021CDJQY-011 and No. 2020CDJQY-Z003), and Chongqing Natural Science Foundation (Grants No.cstc2020jcyj-msxmX0814). 
%
%
SC is supported by a Starting Grant from the Belgian Francqui Foundation.
JMC and BL are supported by NSERC (Natural Sciences and Engineering Research Council), Canada. 
GC is funded by Swiss National Science Foundation (Ambizione Grant).
RJ is supported by the grants IFT Centro de Excelencia 
Severo Ochoa SEV-2016-0597, CEX2020-001007-S and by PID2019-110058GB-C22 
funded by MCIN/AEI/10.13039/501100011033 and by ERDF. 
NL is grateful for the support of the Milner Foundation via the Milner Doctoral Fellowship. 
KFL is partially supported by the U.S. Department of Energy grant DE-SC0022345.
MM is partially  supported   by  the Spanish 
MCIN/AEI/10.13039/501100011033 under
the grant PID2020-113701GB-I00, which includes
ERDF  funds  from  the  European  Union. IFAE  is  partially funded by
the CERCA program of the Generalitat de Catalunya.
ALM is a beneficiary of a FSR Incoming Postdoctoral Fellowship.
%
%
%
%
BS is supported in part by NSF grant PHY-2014075.
JS is supported by the National Natural Science Foundation under Grants No. 12025507, No. 12150015, No.12047503; and is also supported
by the Strategic Priority Research Program and Key Research Program of Frontier Science of
the Chinese Academy of Sciences under Grants No. XDB21010200, No. XDB23010000, and No.
ZDBS-LY-7003 and CAS project for Young Scientists in Basic Research YSBR-006.
XS was supported by NSF's NANOGrav Physics Frontier Center (NSF grants PHY-1430284 and PHY-2020265).
RS is supported by the NSF grant
PHY-1914731 and by the US-Israeli BSF Grant 2018236. 
CT acknowledges financial support by the DFG through the ORIGINS cluster of excellence.
DJW was supported by Academy of Finland grant nos. 324882 and 328958. 
KPX is supported by the University of Nebraska-Lincoln. 
%
SYZ is supported by in part by JSPS KAKENHI Grant Number 21F21026.



\section*{Data Availability}
Data sharing not applicable to this article as no datasets were generated or analysed during the current study.

\bibliography{references}


\end{document}